\documentclass[%
 preprint,
 amsmath,amssymb,
 aps,
]{revtex4-2}

\usepackage{graphicx}
\usepackage{dcolumn}
\usepackage{bm,xcolor}
\usepackage{stmaryrd}
\usepackage{ulem}
\usepackage{subfigure}
\newcommand{\del}[1]{}

\begin{document}

\preprint{Preprint}

\title{Diffuse-interface approach to competition between viscous flow and diffusion in pinch-off dynamics }

 \author{Fukeng Huang}
\email{hfkeng@nus.edu.sg}
 \affiliation{ Department of Mathematics, National	University of Singapore, Singapore, 119076 }

 \author{Weizhu Bao}
\email{matbaowz@nus.edu.sg}
 \affiliation{ Department of Mathematics, National	University of Singapore, Singapore, 119076 }

\author{Tiezheng Qian}%

\email{Corresponding author: maqian@ust.hk}
\affiliation{
 Department of Mathematics, The Hong Kong University of Science and Technology\\
 Clear Water Bay, Kowloon, Hong Kong, PRC
}




\date{\today}

\begin{abstract}
The pinch-off dynamics of a liquid thread has been studied through numerical simulations and theoretical analysis.
Occurring at small length scales, the pinch-off dynamics admits similarity solutions that can be classified into
the Stokes regime and the diffusion-dominated regime, with the latter being recently experimentally observed
in aqueous two-phase systems [Phys. Rev. Lett. 123, 134501 (2019)].
Derived by applying Onsager's variational principle, the Cahn-Hilliard-Navier-Stokes model is employed
as a minimal model capable of describing the interfacial motion driven by
not only advection but also diffusion. By analyzing the free energy dissipation mechanisms in the model,
a characteristic length scale is introduced to measure the competition between diffusion and viscous flow
in interfacial motion. This length scale is typically of nanometer scale for systems far from the critical point,
but can approach micrometer scale for aqueous two-phase systems close to the critical point.
The Cahn-Hilliard-Navier-Stokes model is solved by using an accurate and efficient spectral method
in a cylindrical domain with axisymmetry.
Ample numerical examples are presented to show the pinch-off processes in the Stokes regime
and the diffusion-dominated regime respectively.
In particular, the crossover between these two regimes is investigated numerically and analytically
to reveal how the scaling behaviors of similarity solutions are to be qualitatively changed
as the characteristic length scale is inevitably accessed by the pinching neck of the interface.
Discussions are also provided for numerical examples that are performed for the breakup of long liquid filaments
and show qualitatively different phenomena in different scaling regimes.

\end{abstract}

\maketitle


\section{\label{sec:intro} Introduction}

Aqueous two-phase systems, also known as all-aqueous systems, are formed through phase separation of
an aqueous mixture containing incompatible additives above critical concentrations
\cite{ATPS_A,ATPS_B,ATPS_C}.
Recently, it has been experimentally demonstrated that in an aqueous two-phase system,
a clear interface can be formed between two phases, one polymer rich and the other salt rich, and
the interfacial tension can be widely changed across three orders of magnitude
\cite{ATPS_A,ATPS_B,ATPS_C,ATPS1,ATPS2}.
Thermodynamically, an aqueous mixture undergoes phase separation
when the entropic effect that favors a homogeneous solution is weaker
than the enthalpic effect that favors a separation into two immiscible phases.
The phase-separated system can be controlled to approach or move away from the critical point of
miscibility at room temperature, and the interfacial tension of the water-water interface
is significantly decreased as the critical point is approached.
These systems exhibit novel and interesting interfacial dynamics due to the low interfacial tension.
Furthermore, being environmentally friendly and bio-compatible in general,
aqueous two-phase systems have attracted great attention in many areas for various applications.

The purpose of the present work is to investigate the pinch-off dynamics which occurs
at small length scales and admits similarity solutions that lead to the development of singularities
in space and time \cite{Barenblatt,Eggers1997,Eggers2015}.
As a liquid thread pinches off, the evolving interface exhibits self-similarity
close to the singularity because the length and time scales become much smaller than
those exhibited by the initial interface,
i.e., similarity solutions emerge in the proximity of the singularity
since there are no imposed or specified length or time scales.
In addition, the emergence of similarity solutions requires intrinsic length scales,
determined by liquid properties, to be either regarded as zero or taken as infinity.
For decades there have been numerous experimental, theoretical, and computational works
dedicated to the study of pinch-off dynamics \cite{Eggers1997,Eggers2015}.
However, to the best of our knowledge, most of the existing works were performed in
the regime where the interfacial motion is driven by mass flow.
Therefore, the role of diffusive transport in the pinch-off of liquid threads
largely remains to be explored and studied.
For this purpose, the pinching neck must be sufficiently narrow so that diffusion can
dominate the pinch-off dynamics at small length scales, where viscous flow is suppressed.
Whether such length scales can be experimentally accessed or not depends on liquid properties,
and this makes the aqueous two-phase systems suitable for the study of diffusion-dominated pinch-off
\cite{Xu2019}.

It is interesting to note that in the study of contact line dynamics \cite{Huh-Scriven,deGennes1985,Bonn2009},
the role of diffusion has been investigated at small length scales \cite{Vinals2000,Jacqmin2000,Feng2010}.
As a classical problem in continuum fluid dynamics, the moving contact line is incompatible with
the no-slip boundary condition because the latter leads to a non-integrable singularity in viscous stress.
It has been shown that even if the no-slip condition is applied, the stress singularity
at the contact line can be regularized by introducing diffusion between two immiscible fluid phases
\cite{Vinals2000,Jacqmin2000,Feng2010,Qian2003,Qian2004,Qian2006}.
In the study of solid-state dewetting of thin films caused by capillarity-driven surface diffusion
\cite{Thompson2012}, the retraction and pinch-off of edges \cite{MASS_SHEDDING2000} show some similarity to
the pinch-off of a liquid thread.
Here we use the contact line dynamics and solid-state dewetting to exemplify the role of diffusion
in interfacial motion at small length scales.

The aqueous two-phase systems inject new ingredients into the classical problem of pinch-off dynamics
because of the enhanced diffusion and the weakened capillary force and flow close to the critical point.
This can be quantified by introducing a characteristic length scale
determined by the competition between diffusion and viscous flow.
The magnitude of this length scale depends on liquid properties.
It is typically of nanometer scale for systems far from the critical point,
but can approach micrometer scale for aqueous two-phase systems that are located close enough to
the critical point through accurate control.
At micrometer scale, the characteristic length scale becomes experimentally accessible
in the pinch-off processes, and hence an intrinsic length scale enters into the system.
Theoretically, it is of great interest to investigate the effects of this length scale on
the existence and properties of self-similarity solutions.
Recently, it has been experimentally observed that when the interfacial tension is lower than normal liquids
by two to three orders, the pinch-off dynamics is governed by bulk diffusion \cite{Xu2019},
showing a diffusion-dominated scaling behavior \cite{Aagesen2010} that is distinct from that
dominated by advection in the Stokes regime \cite{Lister1998,Lister1999,Nagel1999}.
The ultralow interfacial tension means that the aqueous two-phase system is controlled to be very
close to the critical point. As a result, the characteristic length scale is at micrometer scale.
As the pinching neck becomes narrower than this length scale, bulk diffusion is less dissipative
than viscous flow, and the diffusion-dominated scaling behavior emerges.
Physically, if the two-phase system is brought closer to the critical point,
then the diffusive transport becomes less dissipative in moving the interface.

We will employ the Cahn-Hilliard-Navier-Stokes (CHNS) model for immiscible two-phase fluids
to investigate the pinch-off dynamics. With the diffuse interface stabilized by a free energy functional,
this phase-field model is a minimal model that incorporates various mechanisms with thermodynamic consistency,
capable of capturing the interfacial motion driven by not only advection but also diffusion,
a salient feature of interfacial dynamics close to the critical point.
The focus will be on the competition between diffusive dissipation and viscous dissipation,
and special attention will be paid to the crossover between the Stokes regime and the diffusion-dominated regime.
Analytical and numerical results of our study will reveal how the singular behaviors of pinch-off dynamics
are to be qualitatively changed as the characteristic length scale is inevitably accessed
by the pinching neck of the interface.
Although the present study has limitations in that the two fluid phases separated by the interface
have equal density, equal viscosity, and equal mobility, we have employed the CHNS model
to demonstrate the presence of two distinct scaling regimes and the crossover between them.

This paper is organized as follows.
In Sec. \ref{sec:model}, the CHNS model is derived by applying Onsager's variational principle
\cite{Onsager1931a, Onsager1931b,Qian2006, Doi2011, Doi2013},
and the characteristic length scale is determined by considering
the competition between diffusion and viscous flow.
The dimensionless equation system is presented with important dimensionless parameters, and
the simulated systems are also described.
In Sec. \ref{sec:results_discussion}, numerical results are presented for the Stokes and diffusion-dominated regimes.
The crossover between the two regimes is also numerically studied to demonstrate
the characteristic length scale in control.
In Sec. \ref{sec:conclusion}, the paper is concluded with a few remarks.

\section{\label{sec:model} Modeling and simulation for two-phase immiscible flows}

\subsection{\label{sec:CHNS} The Cahn-Hilliard-Navier-Stokes model}

Consider a multi-component fluid in which two phases can coexist in equilibrium. Its thermodynamic properties
can be described by using a Ginzberg-Landau-type free energy functional.
We use the Cahn-Hilliard (CH) free energy functional given by \cite{CH1958}
\begin{equation}\label{eq:CH-free-energy}
F_{CH}[\phi (\mathbf{r} )]=\int \left[\displaystyle\frac{K}{2}\left(\nabla\phi \right)^2 +f(\phi) \right] d\mathbf{r},
\end{equation}
in which $\phi:=\phi(\mathbf{r})$ is the phase-field variable to measure the local relative concentration,
$f(\phi)$ is the Helmholtz free energy density for a homogeneous fluid, and $K$ is a positive material parameter.
The free energy density $f$ is given by $f(\phi)=-\frac{\alpha}{2}\phi^2 + \frac{\beta}{4}\phi^4$,
which has a double-well structure to stabilize the fluid-fluid interface between the two stable phases
$\phi_{\pm}=\pm \phi_0=\pm \sqrt{ \frac{\alpha}{\beta} }$, with $\alpha$ and $\beta$ being two positive parameters.
Subject to appropriate boundary conditions, $F_{CH}[\phi (\mathbf{r} )]$ can be minimized
to stabilize a flat interface between the two equilibrium phases of $\phi=\pm \phi_0$.
This gives the characteristic length scale $\xi=\sqrt{\frac{K}{\alpha}}$
for the interfacial thickness and the interfacial tension $\gamma=\frac{2\sqrt{2}\alpha^2\xi}{3\beta}$.
Note that in many literatures, $\phi_0$ is made to equal $1$ through a rescaling. Here we deliberately keep
$\phi_0=\sqrt{ \frac{\alpha}{\beta} }$, which measures the distance away from the critical point.

For an incompressible fluid, the velocity field $\mathbf{v}$ satisfies the incompressibility condition
$\nabla\cdot \mathbf{v}=0$, and the phase field $\phi$ satisfies the continuity equation
\begin{equation}\label{eq:continuity-phi}
\displaystyle\frac{\partial \phi}{\partial t}=-\nabla\cdot\mathbf{J}
=-\nabla\cdot\left( \phi\mathbf{v} + \mathbf{j} \right),
\end{equation}
where $\mathbf{J}=\mathbf{\phi\mathbf{v} + \mathbf{j}}$ is the total current density,
in which $\phi\mathbf{v}$ is caused by flow and $\mathbf{j}$ is the diffusive current density.

Hydrodynamic equations for immiscible two-phase flows can be derived by
applying Onsager's variational principle, which is outlined in Appendix \ref{sec:Onsager}.
For this purpose, we need to find the Rayleighian ${\cal {R}}=\dot{F}_{CH}+\Phi$
in the bulk region where the fluid-fluid interfaces reside in the diffuse-interface modeling.
Here $\dot{F}_{CH}$ is the rate of change of $F_{CH}[\phi (\mathbf{r} )]$, given by
\begin{equation}\label{eq:CH-free-energy-rate}
\dot{F}_{CH}[\phi (\mathbf{r} )]=\int \mu\displaystyle\frac{\partial \phi}{\partial t} d\mathbf{r}
=\int  \nabla\mu\cdot\left( \mathbf{\phi\mathbf{v} + \mathbf{j}} \right) d\mathbf{r},
\end{equation}
in which $\mu\equiv \frac{\delta F_{CH} }{\delta\phi}$ is the chemical potential, given by
$\mu=-K\nabla^2\phi + \frac {df(\phi)}{d\phi}$, and the continuity equation (\ref{eq:continuity-phi}) has been used
with the impermeability conditions for $\mathbf{v}$ and $\mathbf{j}$ at the solid boundary.
The other part in ${\cal {R}}$ is the dissipation functional $\Phi$, given by
\begin{equation}\label{eq:dissipation-functional-B}
\Phi_B=\int  \displaystyle\frac{\eta}{4}
\left[ \nabla\mathbf{v} + \left( \nabla\mathbf{v} \right)^T \right]^2 d\mathbf{r}
+ \int  \displaystyle\frac{\mathbf{j}^2}{2M} d\mathbf{r},
\end{equation}
in which the first term on the right-hand side is caused by the viscous dissipation
with $\eta$ being the shear viscosity, and the second term is caused by the diffusive dissipation
with $M$ being the mobility coefficient.

Subject to the incompressibility condition, the Rayleighian is to be minimized with respect to the rates
$\mathbf{v}$ and $\mathbf{j}$. This gives the force balance equation for $\mathbf{v}$ and
the constitutive equation for $\mathbf{j}$:
\begin{equation}\label{eq:force-balance-equation}
-\nabla p + \nabla\cdot \bm{\sigma}_{\rm visc} - \phi\nabla \mu=0,
\end{equation}
\begin{equation}\label{eq:diffusive-current-density}
\mathbf{j}=-M\nabla\mu,
\end{equation}
in which $p$ is the pressure,
which is the Lagrange multiplier to locally impose $\nabla\cdot \mathbf{v}=0$,
$\bm{\sigma}_{\rm visc}$ is the the Newtonian stress tensor given by
$\bm{\sigma}_{\rm visc}=\eta \left[ \nabla\mathbf{v} + \left( \nabla\mathbf{v} \right)^T \right]$.
Equation (\ref{eq:force-balance-equation}) is the Stokes equation with the capillary force, and
it can be readily generalized to the Navier–Stokes equation
\begin{equation}\label{eq:NS-equation}
\rho\left[ \displaystyle\frac{\partial \mathbf{v}}{\partial t} +
\left(\mathbf{v}\cdot\nabla\right) \mathbf{v}\right]=
-\nabla p + \nabla\cdot \bm{\sigma}_{\rm visc} - \phi\nabla \mu.
\end{equation}
Combining equations (\ref{eq:continuity-phi}) and (\ref{eq:diffusive-current-density}) gives
the advection–diffusion equation for the phase field $\phi$:
\begin{equation}\label{eq:CH-equation}
\displaystyle\frac{\partial \phi}{\partial t} + \mathbf{v}\cdot \nabla \phi
=-\nabla\cdot \mathbf{j} = M \nabla^2 \mu,
\end{equation}
which is the CH equation for a constant $M$.
Equations (\ref{eq:NS-equation}) and (\ref{eq:CH-equation}) are the hydrodynamic equations for
immiscible two-phase flows.

\subsection{\label{sec:length_scale} A characteristic length scale
determined by the competition between diffusion and viscous flow}

A salient feature of the CHNS model for immiscible two-phase flows is the competition
between advection and diffusion. Consider an interface of interfacial velocity $\mathbf{V}_{\rm int}$.
If $\mathbf{V}_{\rm int}$ is realized through diffusion, then the diffusive current density $\mathbf{j}$
is of the order of magnitude of $\phi_0 \mathbf{V}_{\rm int}$ and the dissipation rate so incurred is
about $\frac{1}{M}\left( \phi_0 \mathbf{V}_{\rm int} \right)^2$ per unit volume.
On the other hand, if $\mathbf{V}_{\rm int}$ is realized through advection,
then the dissipation rate so incurred is about
$\eta \left[ \nabla\mathbf{V}_{\rm int} + \left( \nabla\mathbf{V}_{\rm int} \right)^T \right]^2$ per unit volume.
Comparing the above two rates of dissipation, we can define a characteristic length scale
$l_c=\sqrt{\frac{M\eta}{\phi_0^2}}$ at which the rate of diffusive dissipation is comparable to
that of viscous dissipation.
It is worth noting that this length scale has been discussed in a few earlier studies of
diffuse-interface modeling for the contact line dynamics \cite{Jacqmin2000,Feng2010}.
Physically, diffusion is less dissipative than advection over a length scale smaller than $l_c$,
whereas advection is less dissipative than diffusion over a length scale greater than $l_c$.
Based on Onsager's variational principle, the transport mechanism that is less dissipative would
dominate the interfacial motion.
It follows that interfacial motion is dominated by diffusion at small length scales,
but by advection at large length scales.

As the critical point is approached with $\alpha\to 0$,
the interfacial tension $\gamma$ decreases as $\gamma\propto \alpha^{3/2}$, and
the length scale $l_c$ increases as $l_c \propto \alpha^{-1}$ in the mean field theory approximation.
This is obtained by using $\phi_0^2 \propto \alpha$, taking the viscosity $\eta$ as a constant,
and treating the diffusion coefficient $D=2M\alpha$ as a material parameter independent of $\alpha$.
Here $D=2M\alpha$ is derived by approximating the CH equation (\ref{eq:CH-equation})
with the diffusion equation
$\displaystyle\frac{\partial \phi}{\partial t} + \mathbf{v}\cdot \nabla \phi =D \nabla^2 \phi$
for $\phi$ close to $\pm\phi_0$ far away from the diffuse interface.
Physically, $D$ can be expressed using the Einstein relation
$D=\frac{k_BT}{6\pi\eta_s a}$, where $\eta_s$ is the viscosity of the solvent, and $a$ measures
the size of the solute particle, e.g., the radius of gyration of a polymer chain \cite{Xu2019}.
An estimation of $D$ will be given in Sec. \ref{sec:dimensionless}.
By adjusting the ratio of different components, the system can approach or move away
from the critical point, and this can be modelled by varying the parameter $\alpha$.
It is worth emphasizing that the mobility coefficient $M=\frac{D}{2\alpha}$
should not be treated as an invariant while $\alpha$ is varied.

A typical problem that involves interfacial motion at small length scales is the breakup of a liquid thread,
which is called the pinch-off.
While this problem has been extensively studied theoretically, experimentally and numerically,
the crossover from advection-dominated to diffusion-dominated regime has not been systematically investigated.
This is mostly due to the fact that the characteristic length scale $l_c$ is typically too small
(of nanometer scale) to be accessed by the evolving interfaces in experiments.
As the critical point is approached, $l_c\propto \alpha^{-1}$ can be made sufficiently large,
and the diffusion-dominated regime becomes experimentally accessible at micrometer scale \cite{Xu2019}.
In order for the crossover to occur at micrometer scale, accurate control of immiscible two-phase systems
is needed to locate them close enough to the critical point.
A phase-field method with volume preserving Allen–Cahn type phase equation has been used to
simulate the retraction and pinch-off of a liquid filament \cite{yang2006JCP}.
In the present work, we make use of the CH equation (\ref{eq:CH-equation}) to incorporate
the effect of bulk diffusion on interfacial evolution.

\subsection{\label{sec:dimensionless} Dimensionless equations and simulated systems }

Numerical simulations are carried out by solving the CHNS system:
\begin{subequations}
\begin{align}
& \frac{\partial \phi}{\partial t}+\mathbf{v} \cdot \nabla \phi=M \nabla^2 \mu,\\
& \mu=-K\nabla^2 \phi-\alpha \phi+\beta \phi^3,\\
& \rho \big(\frac{\partial \mathbf{v} }{\partial t}+\mathbf{v} \cdot \nabla \mathbf{v}  \big)=-\nabla p+\eta \nabla^2 \mathbf{v} +\mu \nabla \phi, \label{eq:CHNS3}\\
& \nabla \cdot \mathbf{v} =0,
\end{align}
\end{subequations}
in a cylindrical domain $\Omega=\{(x,y,z): x^2+y^2 <L^2, z \in (-H,H) \}$.
Here $M$, $K$, $\alpha$, $\beta$, $\rho$, and $\eta$ are material parameters introduced in Sec. \ref{sec:CHNS}.
Note that the pressure $p$ in equation \eqref{eq:CHNS3} is different from that in equation \eqref{eq:NS-equation}
since $-\phi \nabla \mu$ there is replaced by $\mu \nabla \phi$ here.
The boundary conditions on $x^2+y^2=L^2$ are
\begin{equation}
\frac{\partial \phi}{\partial \mathbf{n}}=\frac{\partial \mu}{\partial \mathbf{n}}=\mathbf{v} =0,
\end{equation}
and periodic boundary conditions are applied to $\phi$, $\mu$, and $\mathbf{v}$ on $z=\pm H$.

To nondimensionalize the above system, we use the radius $L$ of the computational domain $\Omega$ as
the length unit and define the following quantities:
\begin{itemize}
\item  $\bar{H}=\frac{H}{L}$, with the dimensionless height of the computational domain being $2\bar{H}$,
\item $\phi_0=\sqrt{\frac{\alpha}{\beta}}$, with the two equilibrium phases separated by a flat interface being of $\phi=\pm \phi_0$,
\item $\bar{\xi}=\frac{\xi}{L} =\frac{1}{L}\sqrt{\frac{K}{\alpha}}$ as the dimensionless interfacial thickness of the diffuse interface,
\item $D=2 M \alpha$ as the diffusion coefficient for $\phi$ close to $\pm\phi_0$ far away from the interface,
\item $l_c=\sqrt{\frac{M\eta}{\phi_0^2}}$ as the characteristic length scale determined from the competition between diffusion and viscous flow,
\item $p_0=\alpha \phi_0^2$ as the unit of pressure,
\item $\tau=\frac{L^2}{D}$ as the unit of time,
\item $A=\frac{\rho D}{\eta}$, which is the ratio of the diffusion coefficient $D$ to the kinematic viscosity $\frac{\eta}{\rho}$,
\item $B=\frac{\eta D}{\alpha \phi_0^2 L^2}=\frac{2 l_c^2}{L^2}$, which is a dimensionless parameter measuring
the characteristic length scale $l_c$ with respect to $L$.
\end{itemize}
Dimensionless variables, denoted by overbar, are defined through the following relations:
\begin{itemize}
\item
$\phi=\phi_0 \bar{\phi}$; \quad $\bullet$
$\frac{\partial}{\partial t}=\frac{1}{\tau} \frac{\partial}{\partial \bar{t}}$; \quad $\bullet$
$\mathbf{v} =\frac{D}{L} \bar{\mathbf{v}} $;\quad $\bullet$
$\nabla =\frac{1}{L} \bar{\nabla}$; \quad $\bullet$
$\mu=\alpha \phi_0 \bar{\mu}$; \quad $\bullet$
$p=p_0\bar{p}$.
\end{itemize}
Using the above definitions, we obtain the dimensionless CHNS system in the cylindrical domain $\bar{\Omega}=\{(\bar{x},\bar{y},\bar{z}): \bar{x}^2+\bar{y}^2 <1, \bar{z} \in (-\bar{H},\bar{H}) \}$ as
\begin{subequations}\label{CHNS}
\begin{align}
& \frac{\partial \bar \phi}{\partial \bar t}+\bar {\mathbf{v} } \cdot \bar{\nabla} \bar \phi=\frac{1}{2} \bar{\nabla}^2 \bar \mu,\label{CHNS1}\\
& \bar \mu=-\bar{\xi}^2 \bar \nabla^2 \bar \phi-\bar \phi+\bar \phi^3,\label{CHNS2}\\
& A B \big(\frac{\partial \bar {\mathbf{v} }}{\partial \bar t}+\bar {\mathbf{v} } \cdot \bar \nabla \bar {\mathbf{v} } \big)=-\bar \nabla \bar p+B \bar \nabla^2 \bar {\mathbf{v} }+\bar \mu \bar \nabla \bar \phi,\label{CHNS3} \\
& \bar \nabla \cdot \bar {\mathbf{v} }=0.
\end{align}
\end{subequations}
The boundary conditions on $\bar{x}^2+\bar{y}^2=1$ are
\begin{equation}
\frac{\partial \bar\phi}{\partial \mathbf{n}}=\frac{\partial \bar \mu}{\partial \mathbf{n}}=\bar {\mathbf{v} }=0,
\end{equation}
and periodic boundary conditions are applied to $\bar \phi$, $\bar \mu$, and $\bar {\mathbf{v} }$ on $\bar{z}=\pm \bar{H}$.

The dimensionless CHNS system involves the dimensionless parameters $\bar \xi$, $A$, and $B$.
The values used for $\bar\xi$ should be small enough ($\bar\xi\le 0.01$) to ensure the accuracy of
diffuse-interface method. However, the computational cost prevents us from using a value for $\bar\xi$ that is too small.
The value used for $A$, which is the inverse of the Schmidt number, is $10^{-4}$ based on the following consideration.
The diffusion coefficient $D$ is estimated by using $D=\frac{k_BT}{6\pi\eta_s a}$, where
$k_B$ is the Boltzmann constant, $T \approx 300\,{\text K}$ is the temperature, $\eta_s$ is the viscosity of the solvent,
and $a\propto \sqrt N$ is the hydrodynamic radius of polymers, with $N$ being the number of polymer units.
Using $\eta_s\approx 1\, {\text {mPa}}\cdot{\text s}$, $N\approx 200$, and $a\approx 2\,{\text {nm}}$,
we have $D\approx 10^{-10}\,{\text m}^2/{\text s}$ \cite{Xu2019}. Combining it with $\eta\approx 1\, {\text {mPa}}\cdot{\text s}$
and $\rho \approx 10^3\,{\text {kg}}/{\text m}^3$, we have $A=\frac{\rho D}{\eta}\approx 10^{-4}$.
For estimation purpose here, we have used $\eta_s \approx \eta \approx 1\, {\text {mPa}}\cdot{\text s}$,
where the viscosity of the solvent $\eta_s$ is approximately equal to the viscosity of the two-phase fluid $\eta$,
and $1\, {\text {mPa}}\cdot{\text s}$ is the dynamic viscosity of water at $20^\circ C$.
The value used for $B$ is adjusted from case to case to modulate the relative importance of diffusion to viscous flow,
with details presented in Sec. \ref{sec:results_discussion}.

From our three-dimensional (3D) simulations, it is noted that given an axisymmetric initial condition in the cylindrical domain,
the axisymmetry can be accurately preserved during the whole dynamic process.
Therefore, in the absence of any evidence for non-axisymmetric breakup modes, we treat the axisymmetric 3D problem
as a reduced two-dimensional (2D) problem by making use of the cylindrical coordinates
to improve the computational efficiency.
Figures \ref{fig:neck_radius_flow}, \ref{fig:mother_satellite_flow}, \ref{fig:neck_radius_flow_a},
and \ref{fig:multiple_drops_diffusion} provide many illustrations for
the simulated systems in cylindrical domain with axisymmetry.
Details of the reduced 2D problem are given in Appendix \ref{sec:numerical}, where a numerical scheme is also presented.

 Finally, we would like to emphasize that to achieve the similarity solutions, the dimenisonless neck radius $\bar{r}_n$
of the pinching neck must be sufficiently small compared to the radius of the computational domain $L=1$.
Furthermore, in the present diffuse-interface model, the interfacial thickness $\bar{\xi}$ must be sufficiently small
compared to the neck radius $\bar{r}_n$, which represents the smallest length scale exhibited by the interfacial shape.
The disparity among $\bar{\xi}$, $\bar{r}_n$, and $L=1$ leads to huge computational cost,
and an efficient and accurate numerical scheme is essential to our simulations.

\section{\label{sec:results_discussion} Results and discussion}

In the pinch-off dynamics, we take the neck radius $r_n$ of the liquid thread as
a characteristic length scale of the interface.
The magnitude of the chemical potential at this length scale is
$\sim \mu_n= \frac{\gamma}{\phi_0 r_n}$, which is obtained from an estimation
of the magnitude of $-K\nabla^2\phi$ in the sharp interface limit
\cite{Vinals2000,Jacqmin2000,Feng2010}.
The diffusive current density $\mathbf{j}=-M\nabla \mu$ is of a magnitude
approximately given by $M\frac{\mu_n}{r_n}\sim \frac{M\gamma}{\phi_0 r_n^2}$,
which may lead to an interfacial velocity of a magnitude
$\sim V_d=\frac{M\gamma}{\phi_0^2 r_n^2}$ in the absence of advection.
On the other hand, in the Stokes regime, balancing the capillary pressure against
the viscous stress leads to an interfacial velocity of a magnitude
$\sim V_\eta =\frac{\gamma}{\eta}$ in the absence of diffusion.
Comparing $V_d$ and $V_\eta$, we obtain
\begin{equation}\label{eq:V_d-V_c}
\displaystyle\frac{V_d}{V_\eta}= \displaystyle\frac{M \eta}{\phi_0^2 r_n^2}
=\displaystyle\frac{l_c^2}{r_n^2},
\end{equation}
in which the length scale $l_c=\sqrt{\frac{M\eta}{\phi_0^2}}$ defined above shows up.
It is readily seen that for $r_n \gg l_c$, we have ${V_\eta} \gg {V_d}$, which means that
the pinch-off dynamics is dominated by viscous flow. This is the Stokes regime, in which
the neck radius is linearly dependent on time, i.e., $r_n (t) \propto V_\eta (t_s-t)$,
with $t_s$ being the time of pinch-off \cite{Nagel1999}.
Numerical results for this regime will be presented in Sec. \ref{sec:Stokes_regime}.
On the other hand, for $r_n \ll l_c$, we have ${V_d} \gg {V_\eta}$, which means that
the pinch-off dynamics is dominated by bulk diffusion.
This is the diffusion-dominated regime, in which the interfacial motion can be effectively
described by the CH equation (\ref{eq:CH-equation}) without $\mathbf{v}\cdot\nabla\phi$.
Numerical results for this regime will be presented in Sec. \ref{sec:diffusion_regime}.
It is also interesting to observe the time evolution of interface which experiences a stage of
$r_n \sim l_c$, at which there is a crossover from the Stokes regime to the diffusion-dominated regime.
Numerical results in this regard will be presented in Sec. \ref{sec:crossover}.

\subsection{\label{sec:Stokes_regime} Stokes regime}

The Stokes regime has been well studied experimentally and theoretically \cite{Lister1998,Lister1999,Nagel1999}.
In this regime, the neck radius of the liquid thread is a linear function of time, and
$V_\eta =\frac{\gamma}{\eta}$ is the only velocity scale.
The internal and external viscosities are equal in our simulations, and hence the simulation results
are compared to $r_n(t)=0.0335 \frac{\gamma}{\eta} (t_s-t)$, which has been obtained from
early numerical simulations \cite{Lister1998,Lister1999} and a scaling theory \cite{Nagel1999}.
Using the dimensionless variables and parameters defined in Sec. \ref{sec:dimensionless}, we have
$\bar{r}_n=0.0316 \frac{\bar{\xi}}{B}\bar{t}^*$, where $\bar{t}^*=\frac{t_s-t}{L^2/D}$ is
the dimensionless time to pinch-off, and $0.0316$ is obtained from the above coefficient value $0.0335$
with $0.0335\times\frac{2\sqrt{2}}{3}=0.0316$.
Our numerical results show quantitative agreement with $\bar{r}_n=0.0316 \frac{\bar{\xi}}{B}\bar{t}^*$,
which has been experimentally confirmed \cite{Nagel1999}.

We have carried out simulations by using small values of $B$ to ensure that the characteristic length scale
$l_c$ is much smaller than the neck radius $r_n$ during its evolution, and hence the pinch-off dynamics is
dominated by viscous flow.
The evolution of the interface is generated by introducing an undulation to the initial interfacial profile,
given by $$r_0(z)=R-\varepsilon \cos\frac{\pi z}{H},$$
where $R$ is the radius of an unperturbed liquid thread,
and $\varepsilon$ is the amplitude of a cosinusoidal undulation, with $r_0 (0)=R-\varepsilon$ and $r_0(\pm H)=R+\varepsilon$.
For small $\varepsilon$, if the undulation wavelength $2H$ exceeds the critical value $2\pi R$,
then the undulation amplitude grows with time, showing an exponential growth in the beginning.

\begin{figure}[!htbp]
 \begin{center}
 \subfigure{ \includegraphics[scale=.32]{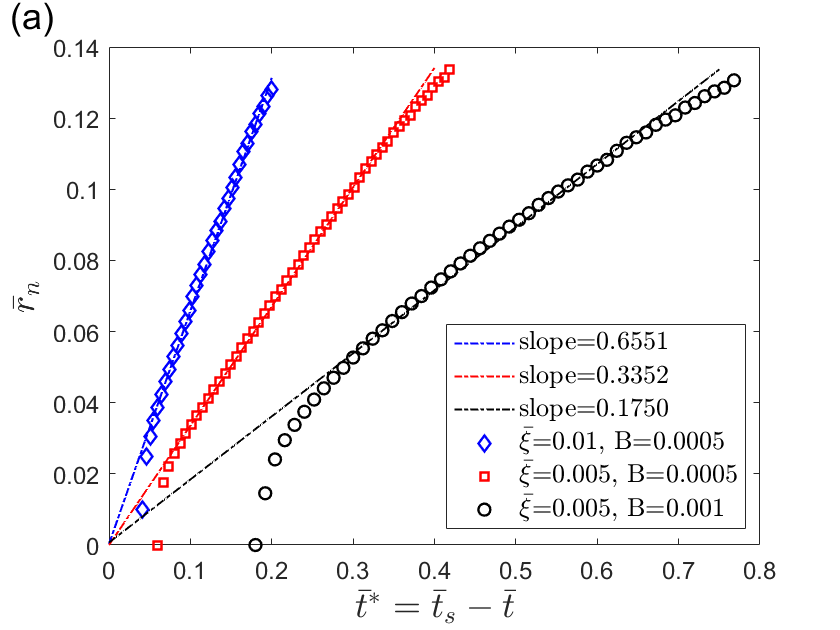}}\\
 \subfigure{ \includegraphics[scale=.28]{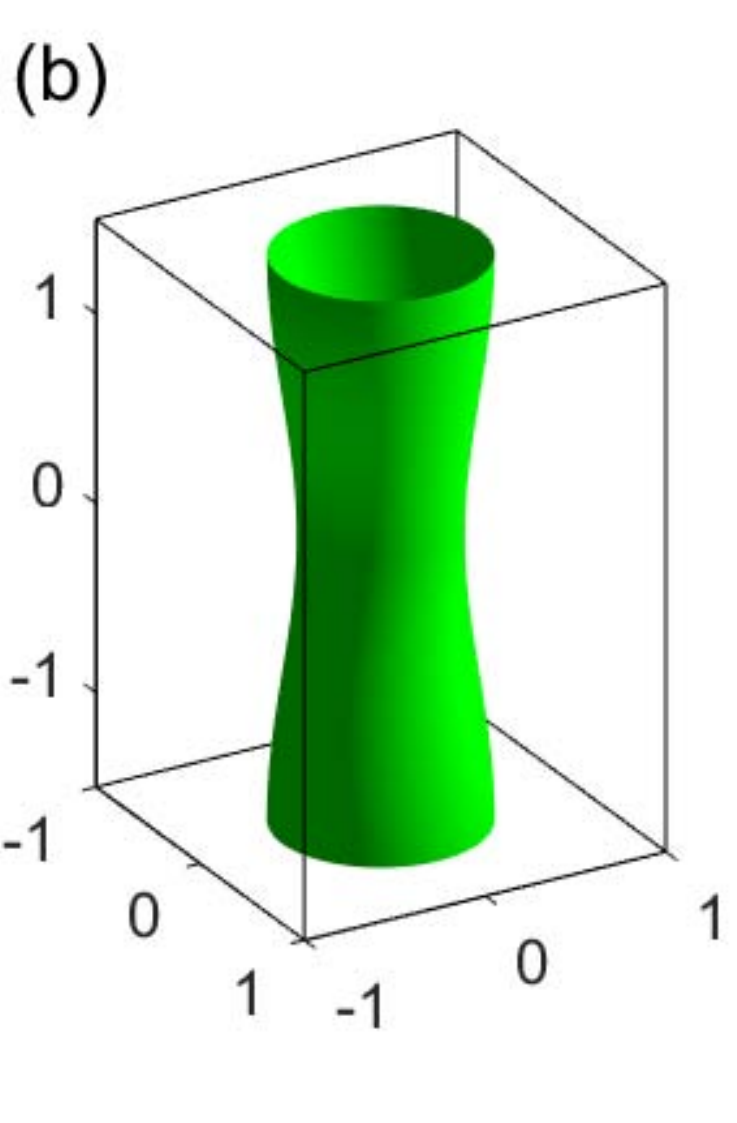}}
 \subfigure{ \includegraphics[scale=.28]{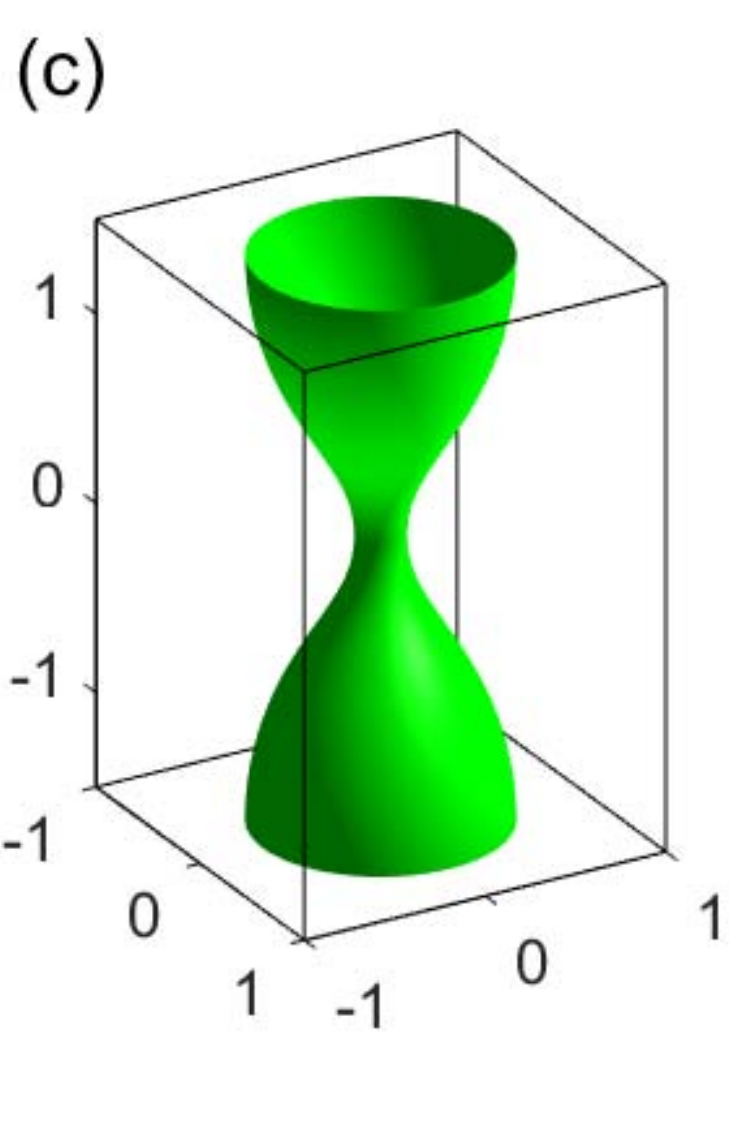}}
 \subfigure{ \includegraphics[scale=.28]{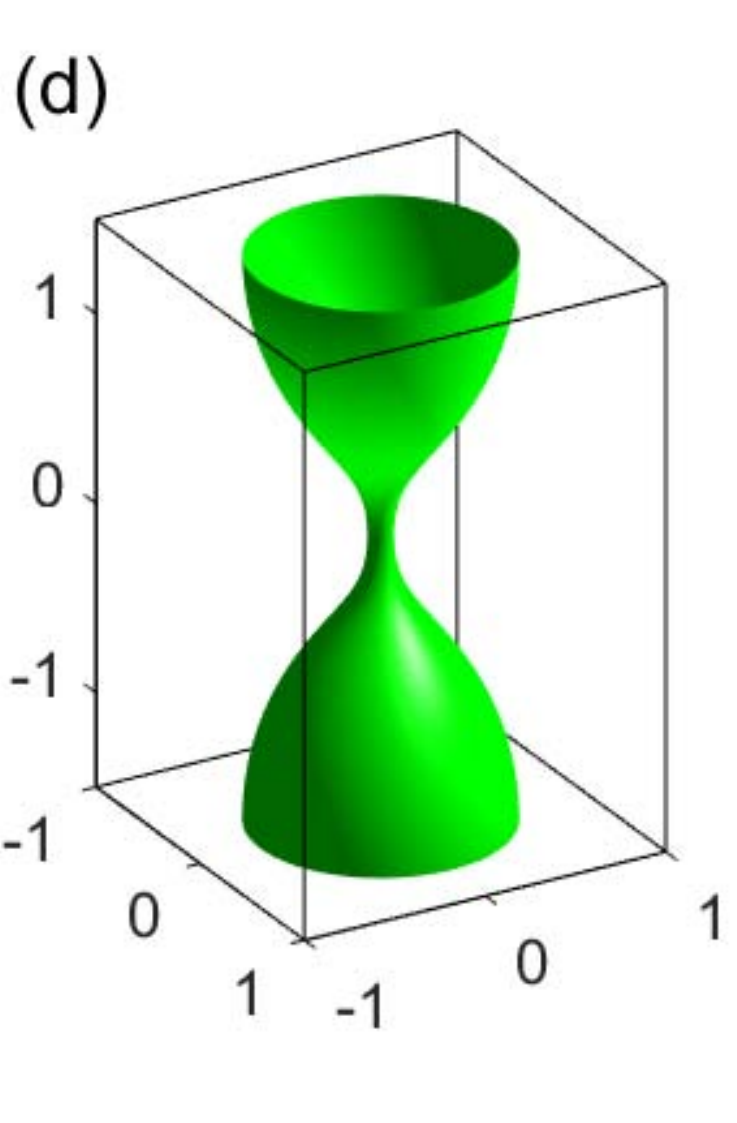}}
 \subfigure{ \includegraphics[scale=.28]{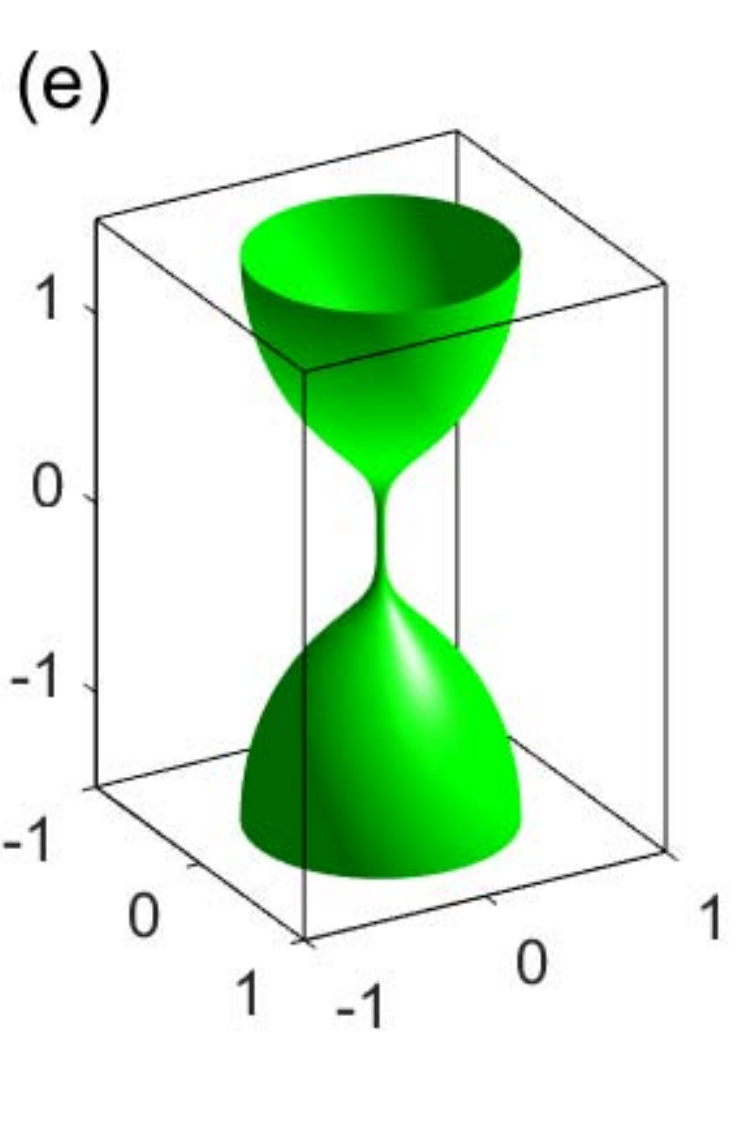}}
 \subfigure{ \includegraphics[scale=.28]{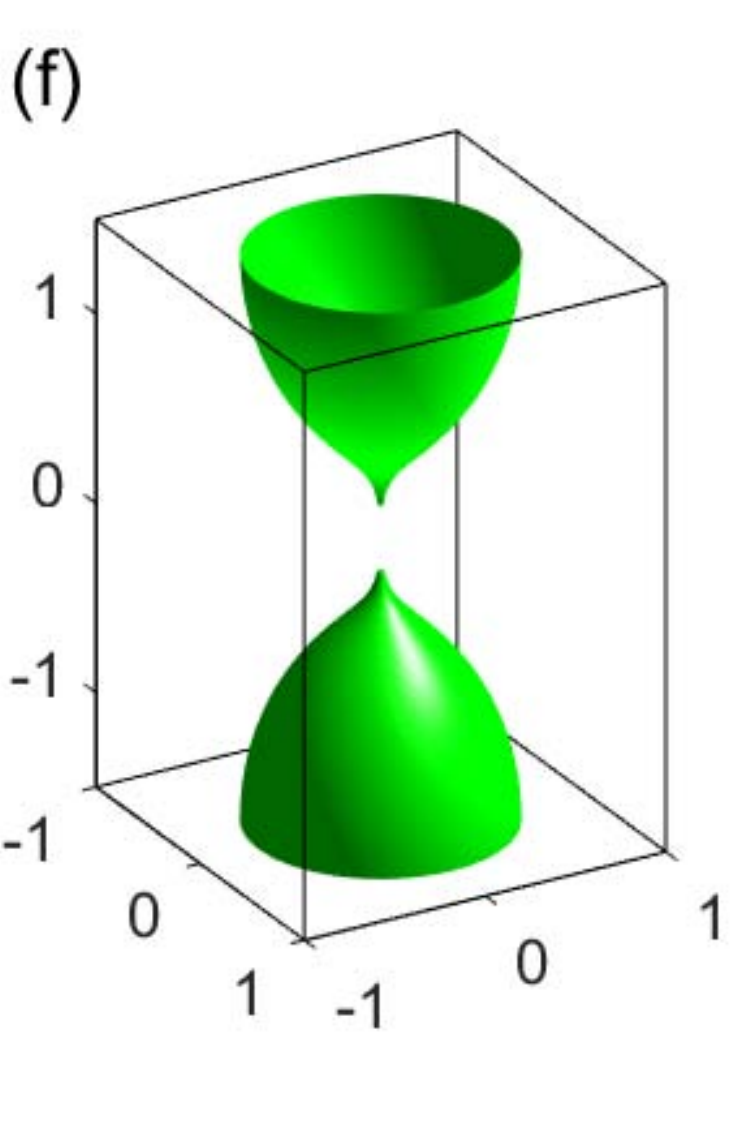}}
 \subfigure{ \includegraphics[scale=.28]{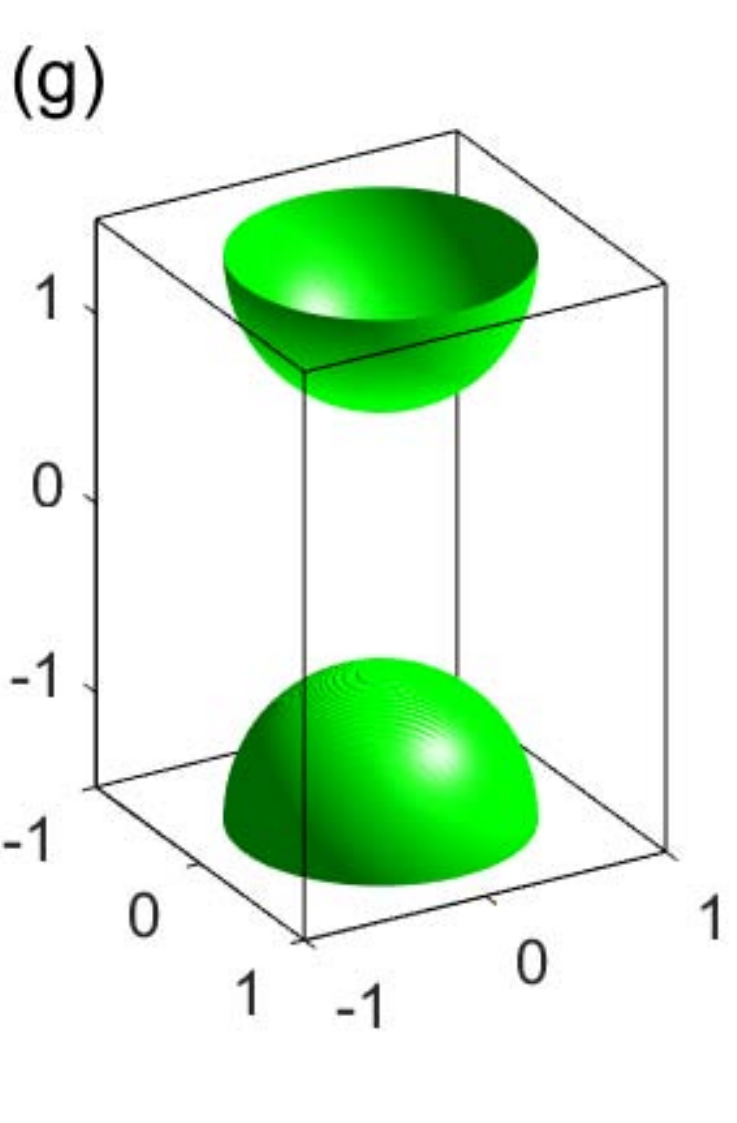}}
 \caption{(a) The dimensionless neck radius $\bar{r}_n$ plotted as a function of
 the dimensionless time to pinch-off $\bar{t}^*=\bar{t}_s-\bar{t}$.
 Using $\bar{H}=1.5$ and the initial interfacial profile with
 $\bar{r}_0(0)=0.4037$ and $\bar{r}_0(\pm \bar{H})=0.5463$,
 numerical results are obtained for three different combinations of $\bar{\xi}$ and $B$.
 The dash-dotted lines are used to indicate the linear dependence, from which
 the slopes are found to be close to $0.0316 \frac{\bar{\xi}}{B}$.
 (b)-(g) are six snapshots taken at $\bar{t}=0.02$, $12.78$, $13.00$, $13.10$, $13.12$, and $14.00$
 for $\bar{\xi}=0.005$ and $B=0.0005$ in (a).
 }\label{fig:neck_radius_flow}
 \end{center}
 \end{figure}

For $2H$ slightly larger than $2\pi R$, the pinch-off occurs at $z=0$ about which the evolving interface is symmetric,
and the neck radius $r_n$ of the liquid thread is measured as a function of time at $z=0$.
Figure \ref{fig:neck_radius_flow} shows the time dependence of $\bar{r}_n$ for different combinations of
$\bar{\xi}$ and $B$.
Note that in the diffuse-interface simulations, there is a lower bound to the physically meaningful range
of $r_n$: it can be as small as a few times of $\xi$, but not smaller.
Close to the pinch-off, the linear dependence of $\bar{r}_n$ on $\bar{t}^*$ shows up,
with the slope being close to $0.0316 \frac{\bar{\xi}}{B}$. The deviation from this analytical value is found
to be small, partly attributed to the finite radius of the cylindrical domain in our computation.
Let the evolving interface have radius $r(z,t)$. It is numerically verified that $r$ and $z$ scale with $t$
as $r \sim z \sim t_s-t$ as $t_s-t\to 0$, exhibiting no preferred direction in space.
All these cases are in the Stokes regime as the Reynolds number $\frac{\rho V_\eta L}{\eta}$,
which can be expressed by $\frac{2\sqrt{2}A\bar\xi}{3B}$, is much smaller than $1$ for $A=10^{-4}$.

\begin{figure}[!htbp]
 \centering
  \subfigure{ \includegraphics[scale=.45]{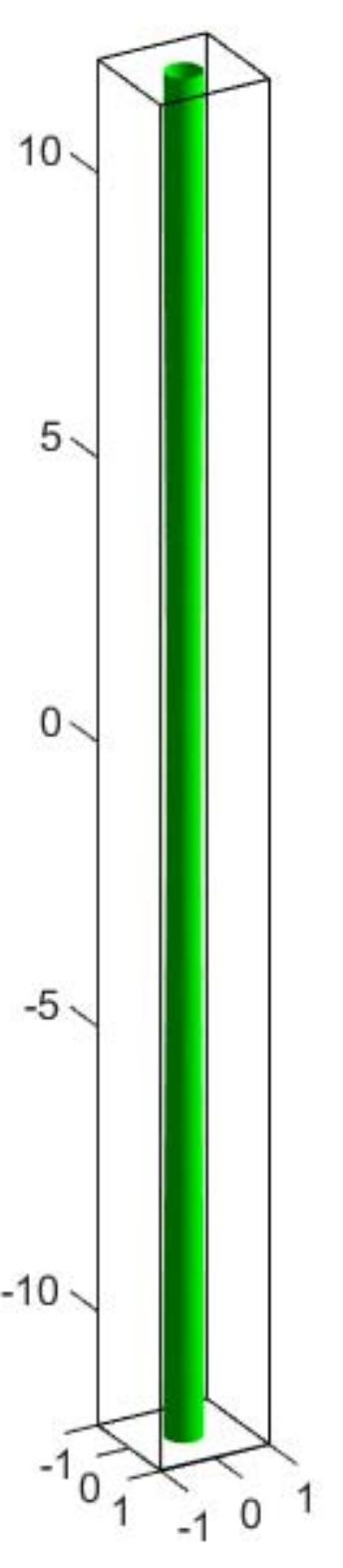}}
  \subfigure{ \includegraphics[scale=.45]{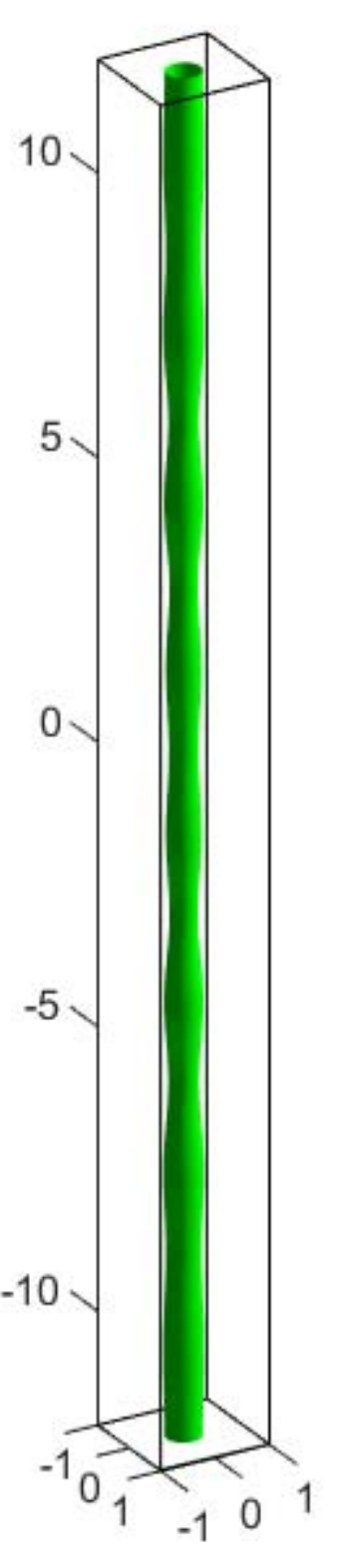}}
  \subfigure{ \includegraphics[scale=.45]{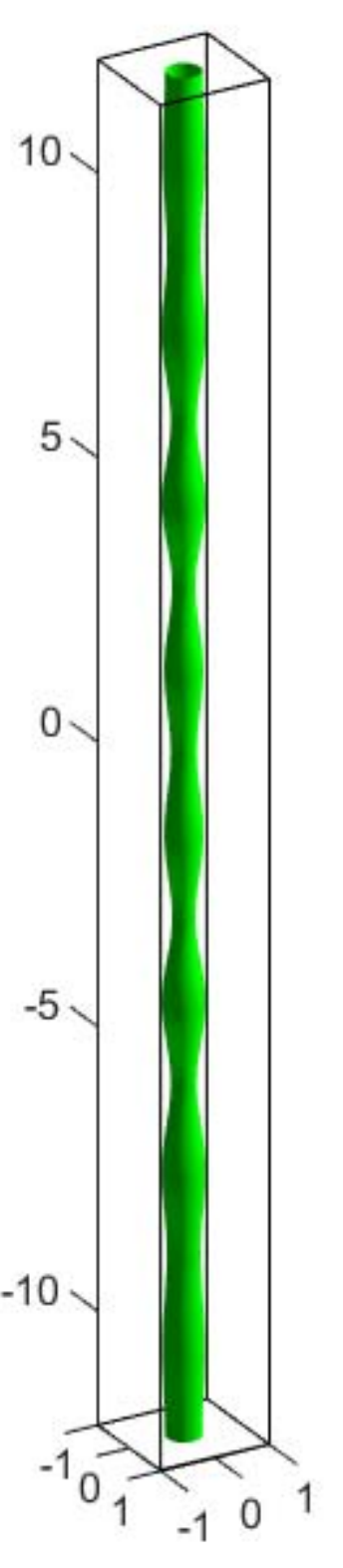}}
  \subfigure{ \includegraphics[scale=.45]{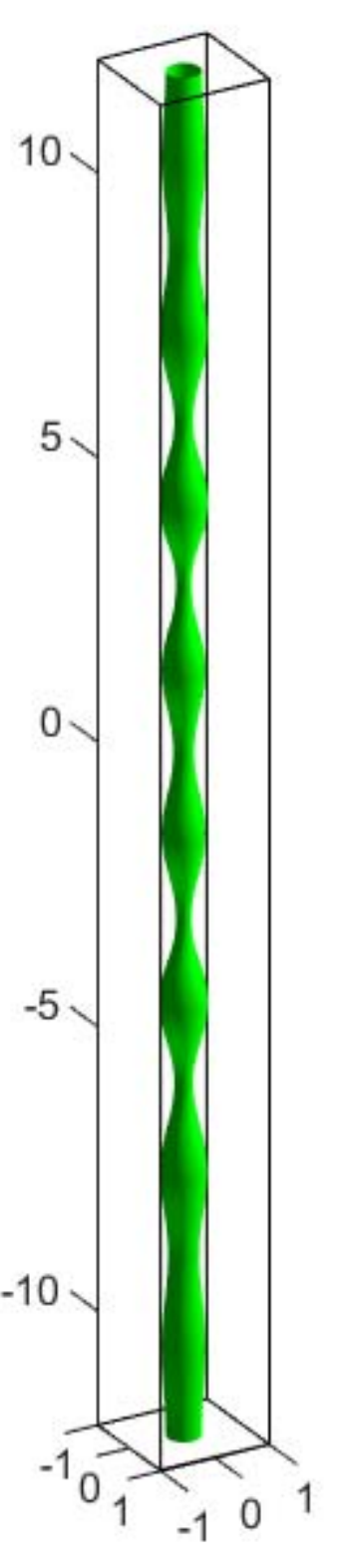}}
  \subfigure{ \includegraphics[scale=.45]{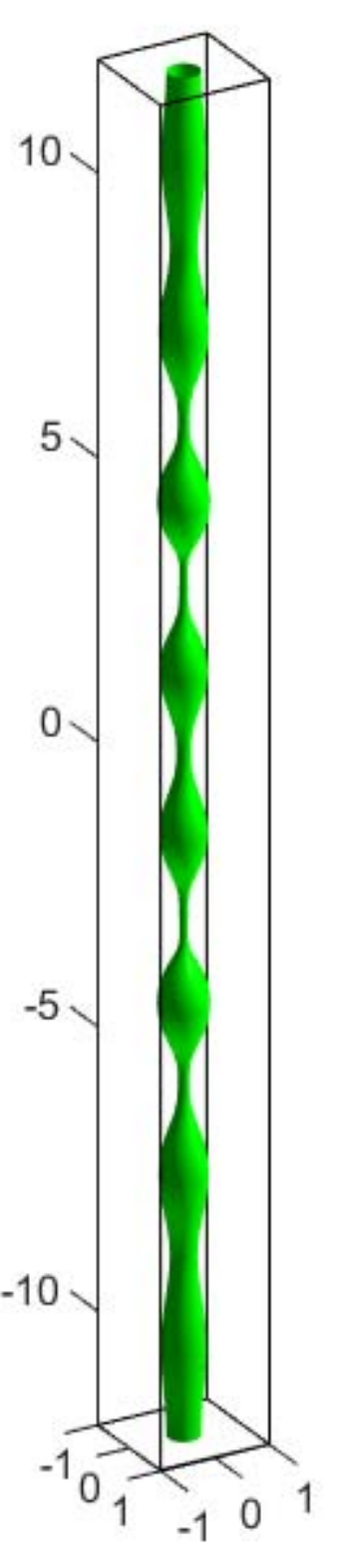}}
  \subfigure{ \includegraphics[scale=.45]{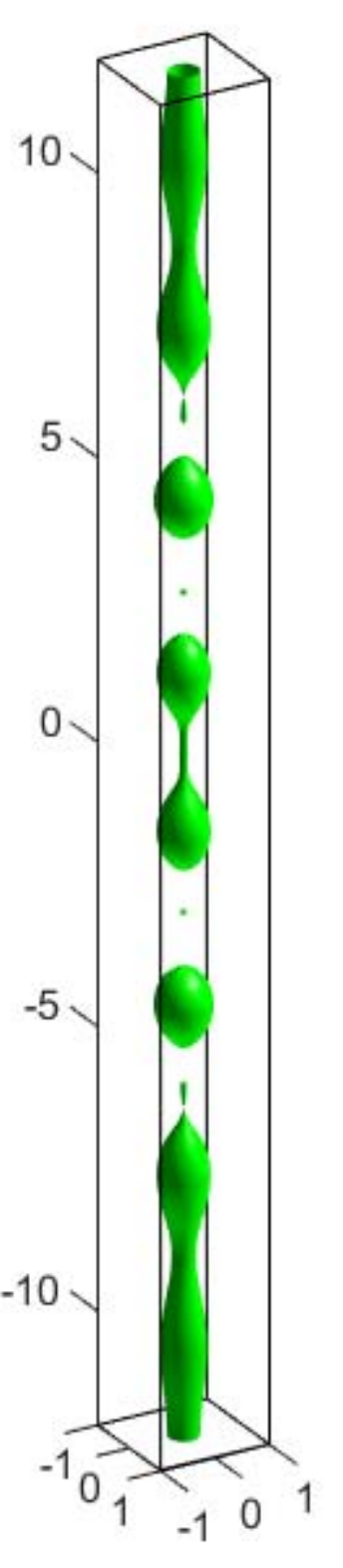}}
  \subfigure{ \includegraphics[scale=.45]{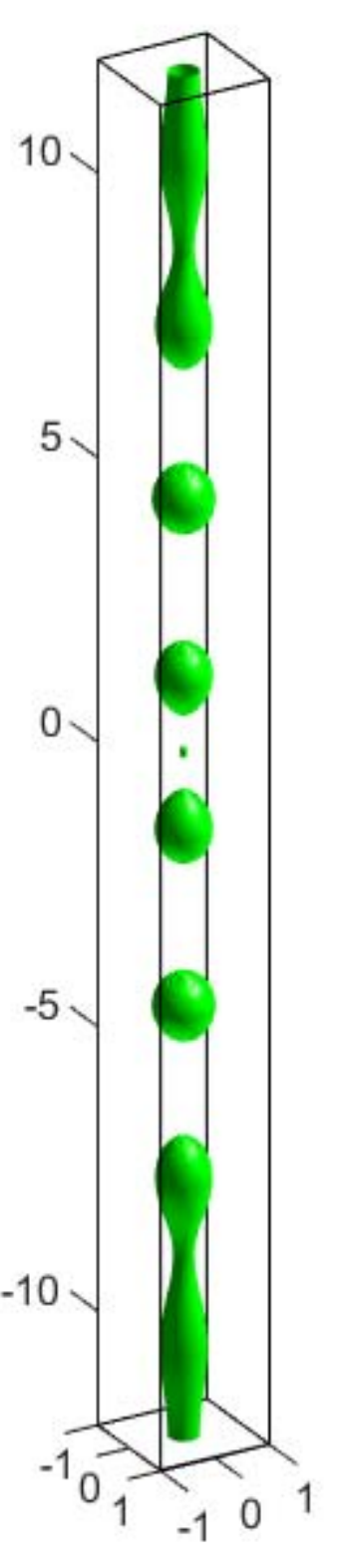}}
  \subfigure{ \includegraphics[scale=.45]{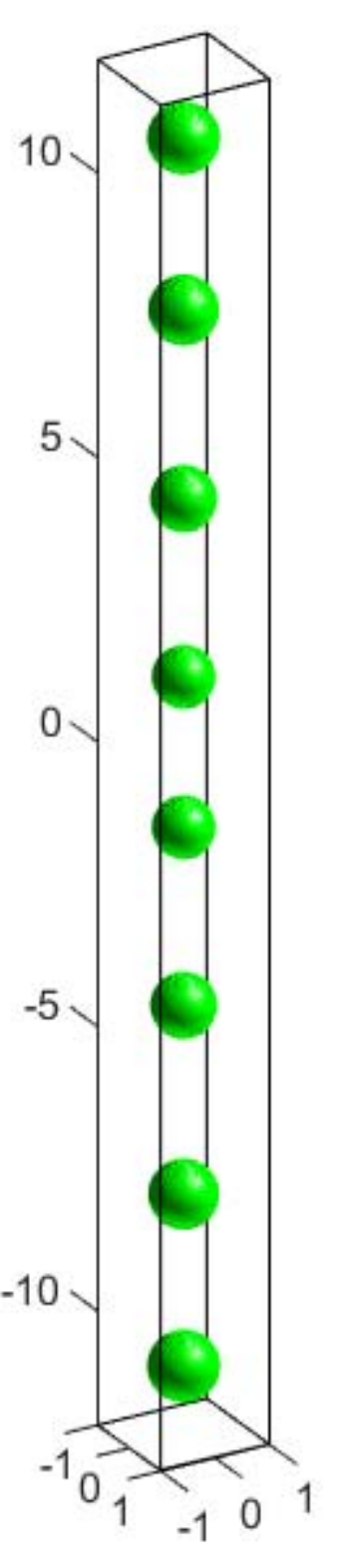}}\\
 \caption{Formation of mother drops and satellite drops in the fragmentation of long liquid filaments.
 Using $\bar{H}=12$ and the initial interfacial profile with
 $\bar{r}_0(0)=0.28$ and $\bar{r}_0(\pm \bar{H})=0.32$,
 numerical results are obtained for $\bar{\xi}=0.02$ and $B=0.0005$.
 The eight snapshots are taken at $\bar{t}=0.008$, $1.64$, $1.8$, $1.92$, $2$, $2.072$, $2.112$, and $3.2$.
 Note that the initial interfacial profile involves only one period of undulation.
 }
 \label{fig:mother_satellite_flow}
 \end{figure}

The investigation of breakup of long liquid filaments
has been a classical problem in fluid mechanics \cite{castrejon2012,driessen2013,schulkes1996}.
Driven by interfacial tension, the fragmentation of a long liquid filament leads to
an array of uniformly spaced large drops, called mother drops,
with smaller droplets in between, called satellite drops \cite{Tjahjadi1992}.
This phenomenon has been observed in our simulations by using $2H$ much larger than $2\pi R$.
Although the initial perturbation in $r_0(z)$ involves only one period of undulation, multiple breakup events
occur in the course of time, leading to mother drops and satellite drops arrayed in the axial direction.
Due to Ostwald ripening inherent in the phase-field simulations \cite{yue2007,fan1998},
small satellite drops can only exist for a short duration of time before they dissolve via diffusion.
As a thermodynamic process, Ostwald ripening spontaneously occurs because larger droplets are more
favored than smaller droplets in free energy.
Figure \ref{fig:mother_satellite_flow} presents a sequence of snapshots showing the time evolution
of a long liquid thread.

It is seen that the development of an instability mode shows several periods and eventually leads to
the production of eight mother drops in total, with small and transient satellites in between.
Note that the first four mother drops nearest to $z=0$
are accurately positioned by the crests of the instability mode.
Figure \ref{fig:mother_satellite_flow} also shows some ``end effect'' that is regarded as an artifact,
because a prescribed length $2\bar{H}$ does not necessarily coincide with an integer multiple of
the dominant wavelength.

\begin{figure}[!htbp]
 \centering
 \subfigure{ \includegraphics[scale=.32]{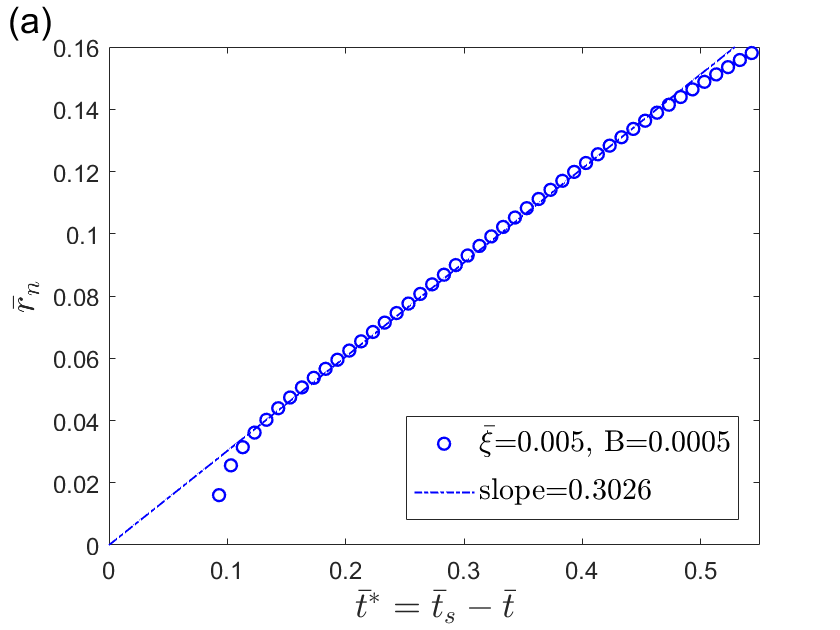}}\\
 \subfigure{ \includegraphics[scale=.28]{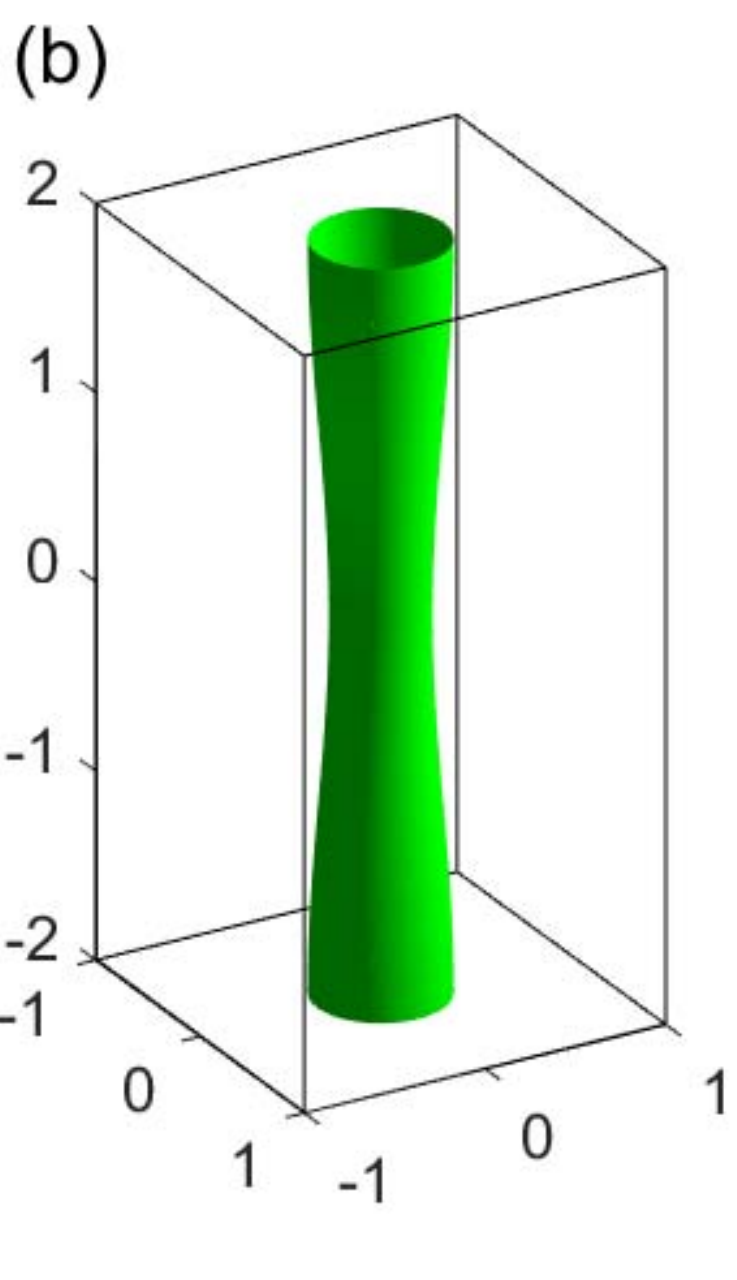}}
 \subfigure{ \includegraphics[scale=.28]{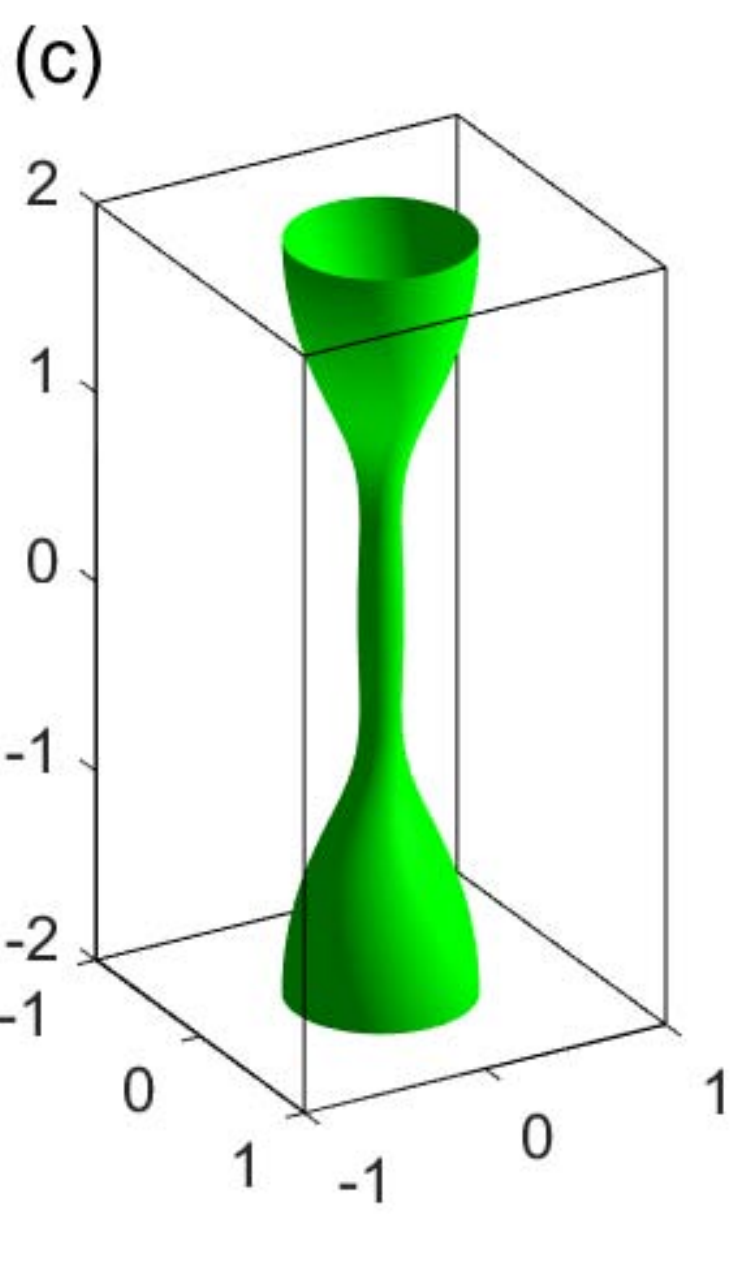}}
 \subfigure{ \includegraphics[scale=.28]{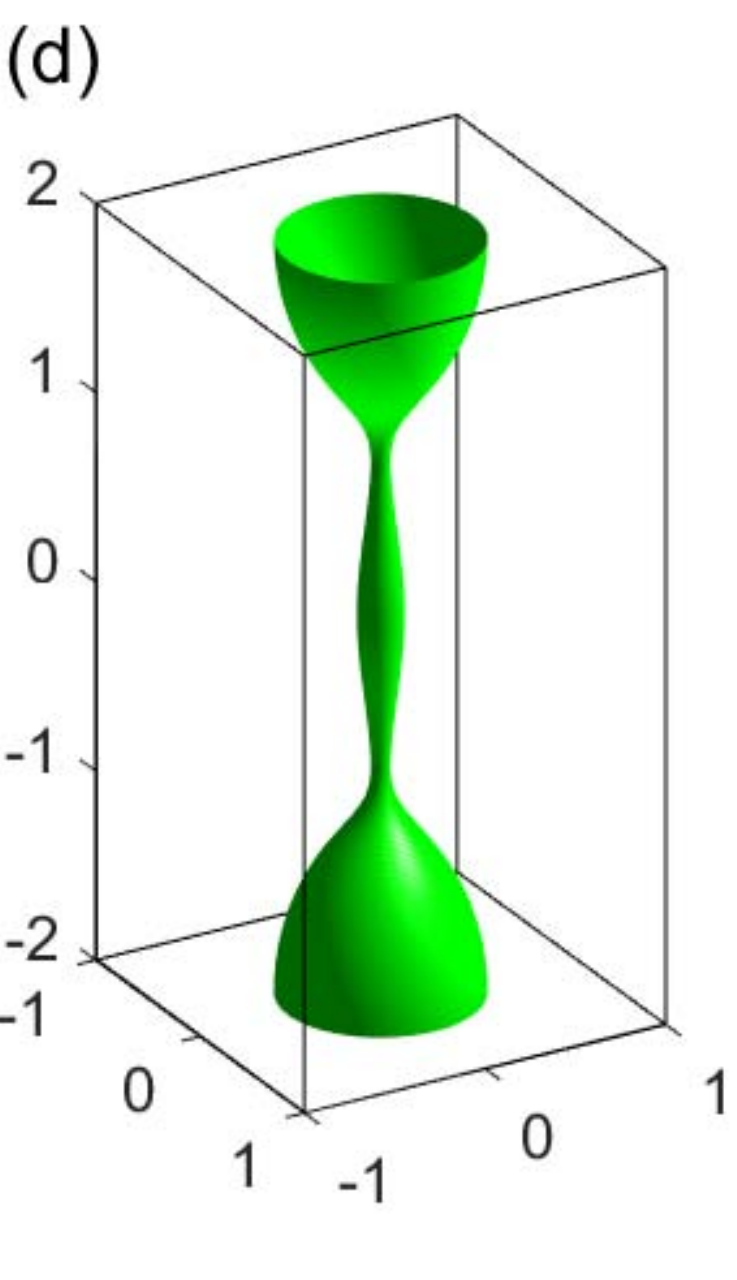}}
 \subfigure{ \includegraphics[scale=.28]{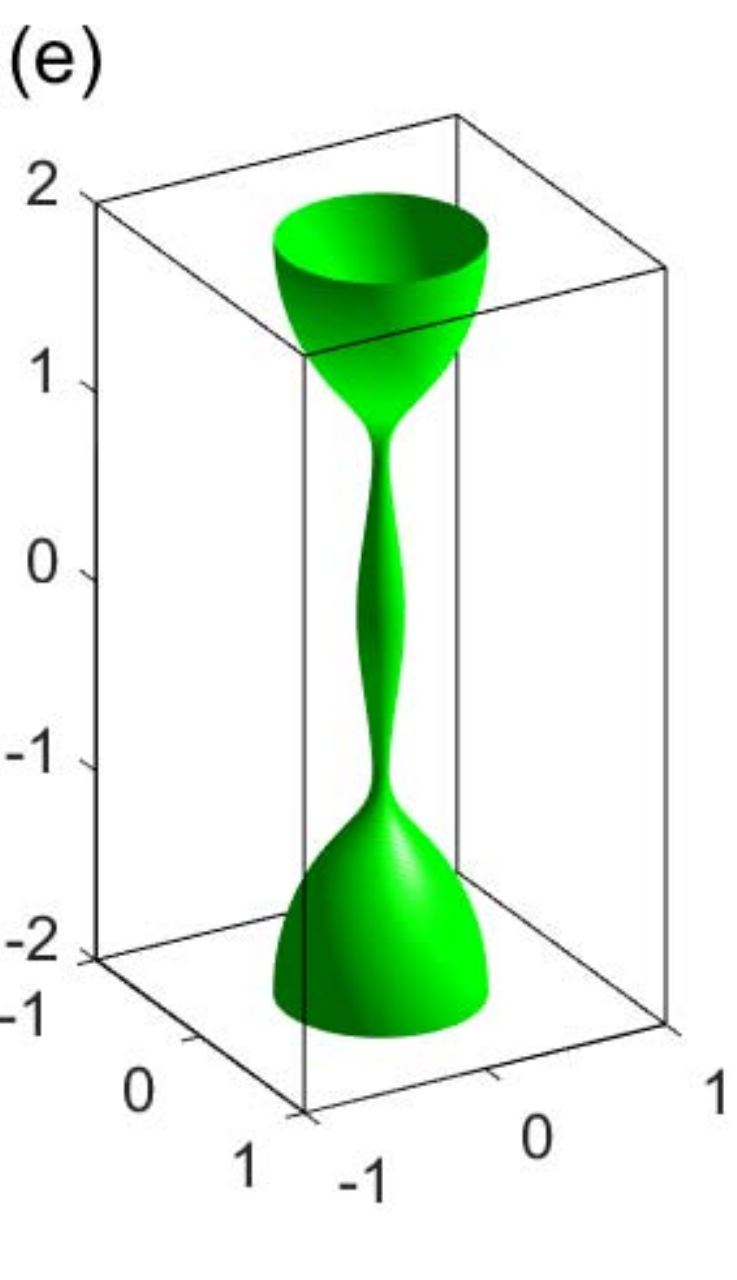}}
 \subfigure{ \includegraphics[scale=.28]{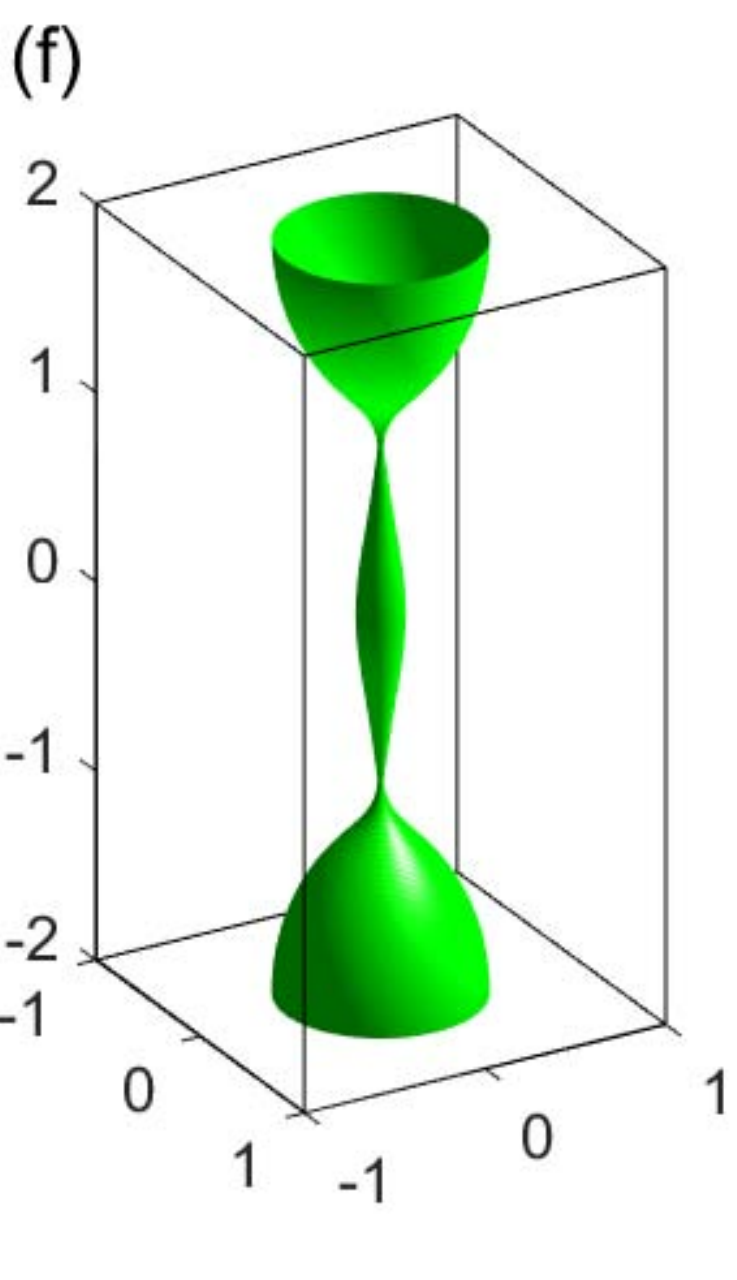}}
 \subfigure{ \includegraphics[scale=.28]{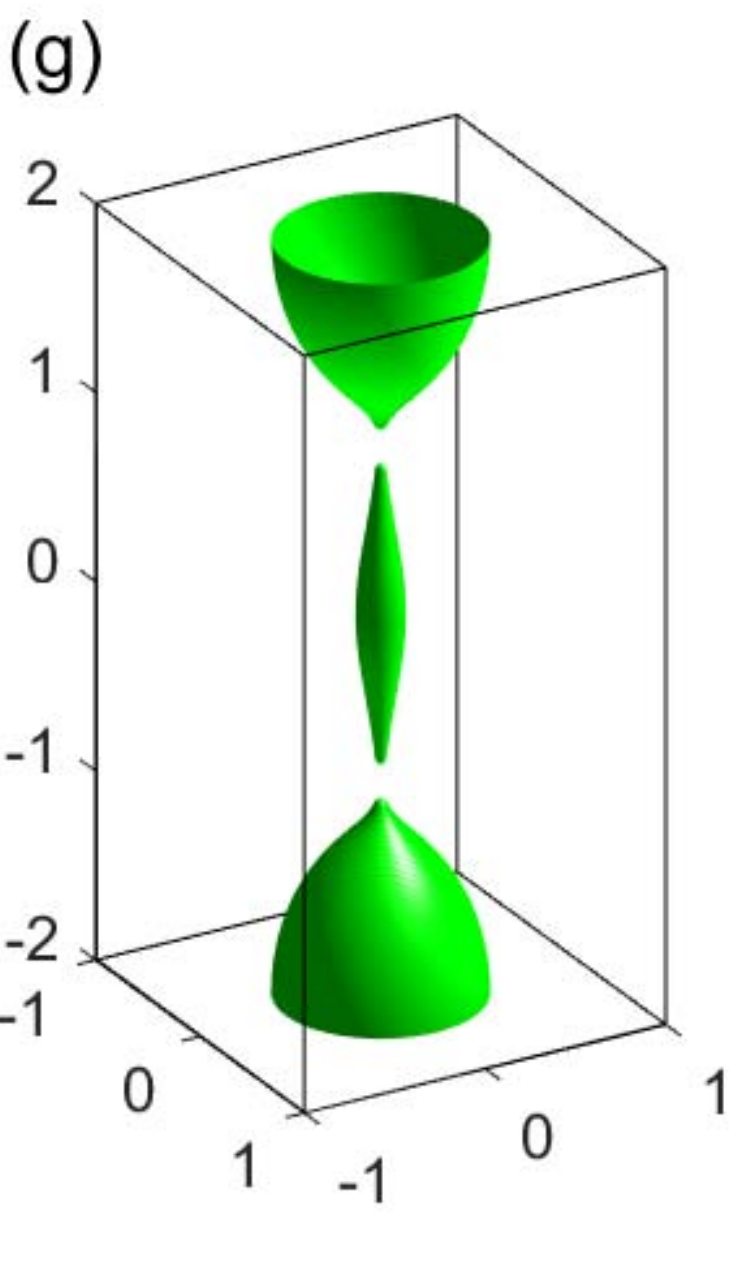}}
 \caption{(a) The dimensionless neck radius $\bar{r}_n$
 plotted as a function of the dimensionless time to pinch-off $\bar{t}^*=\bar{t}_s-\bar{t}$.
 Using $\bar{H}=2$ and the initial interfacial profile with
 $\bar{r}_0(0)=0.25$ and $\bar{r}_0(\pm \bar{H})=0.35$,
 numerical results are obtained for $\bar{\xi}=0.005$ and $B=0.0005$.
 The dash-dotted line is used to indicate the linear dependence, from which
 the slope is found to be close to $0.0316 \frac{\bar{\xi}}{B}$.
 (b)-(g) are six snapshots taken at $\bar{t}=0.02$, $1.80$, $2.06$, $2.08$, $2.12$, and $2.13$.
 Note that the initial interfacial profile involves only one period of undulation.
}
 \label{fig:neck_radius_flow_a}
 \end{figure}
The breakup of long liquid filaments presents many pinch-off processes away from the plane of $z=0$.
Our numerical simulations show that in these processes, the neck radius is still a linear function of time.
Figure \ref{fig:neck_radius_flow_a} shows the time dependence of $\bar{r}_n$, which is measured from
two pinch-off processes that are symmetric about $z=0$ and occur simultaneously to produce a drop centered at $z=0$.
In this case, the interfacial profile $\bar{r}(\bar{z},\bar{t})$ has two equal minima
$\bar{r}_n(\bar{t})$ occurring at $\bar{z}=\pm \bar{z}_n(\bar{t})$,
where $\bar{z}_n(\bar{t})$ slightly varies with time around $\bar{z}=0.84$.
It is interesting to note that for $\bar{r}_n\ge 0.04$, the linear dependence of $\bar{r}_n$ on $\bar{t}^*$
does give a slope close to $0.0316 \frac{\bar{\xi}}{B}$.
Compared to figure \ref{fig:neck_radius_flow} for interfacial profiles {\it symmetric} about
the pinch-off point at $z=0$, the interfacial profiles in figure \ref{fig:neck_radius_flow_a}
are {\it asymmetric} about $\bar{z}=\pm \bar{z}_n(\bar{t})$
where the neck radius $\bar{r}_n(\bar{t})$ is located.
Nevertheless, both figures show that $\bar{r}_n$ linearly depends on $\bar{t}^*$, with the slope
being close to $0.0316 \frac{\bar{\xi}}{B}$.
Note that the asymmetric pinch-off shown in figure \ref{fig:neck_radius_flow_a} displays the salient features
of the self-similar profiles in the Stokes regime \cite{Lister1999,Nagel1999},
characterized by two cones of different opening angles.
More numerical results can be found in Appendix \ref{sec:numerical}.

\subsection{\label{sec:diffusion_regime} Diffusion-dominated regime}

For $r_n \ll l_c$, the pinch-off dynamics is dominated by bulk diffusion with viscous flow being suppressed.
In this regime, a self-similar solution exists near pinch-off.
Scaling $\mathbf{r}$ by ${\cal A}\left(t_s-t\right)^a$ and $\mu$ by ${\cal B}\left(t_s-t\right)^b$, we write
the self-similar solution for $\mu$ as
\begin{equation}\label{eq:self-similar-mu}
\mu={\cal B}\left(t_s-t\right)^b h \left(\displaystyle\frac{\mathbf{r}}{{\cal A}\left(t_s-t\right)^a}\right),
\end{equation}
in which $\mathbf{r}$ is measured relative to the center of the neck,
$t_s$ is the time of pinch-off, ${\cal A}$ and ${\cal B}$ are scaling constants
depending on material properties, and the exponents $a$ and $b$ are determined as follows.
From the characteristic magnitude of the chemical potential $\mu_n= \frac{\gamma}{\phi_0 r_n}$,
we have $b=-a$. Furthermore, from the characteristic magnitude of the interfacial velocity
$V_d=\frac{M\gamma}{\phi_0^2 r_n^2}$ due to diffusion, we have $a-1=-2a$. These scaling arguments
lead to $a=-b=\frac{1}{3}$, which has been experimentally observed
in the diffusion-dominated pinch-off \cite{Aagesen2010,Xu2019}.

We have carried out simulations by using large values of $B$ to ensure that
the characteristic length scale $l_c$ is much larger than the neck radius $r_n$ during its evolution,
and hence the pinch-off dynamics is dominated by diffusion.
The evolution of the interface is still initiated by using the initial interfacial profile $r_0(z)$,
which involves only one period of undulation. For sufficiently large values of $B$,
the pinch-off occurs at $z=0$, unless $2H$ is much larger than $2\pi R$.
In other words, multiple breakup events may still occur in the course of time,
but they require a much larger ratio of $2H$ to $2\pi R$ when compared to
those in the Stokes regime presented in Sec. \ref{sec:Stokes_regime}.
This distinction can be understood by looking at the sequences of multiple breakup events
in the diffusion-dominated regime and noting their difference from those in the Stokes regime.
\begin{figure}[!htbp]
 \centering
  \subfigure{ \includegraphics[scale=.32]{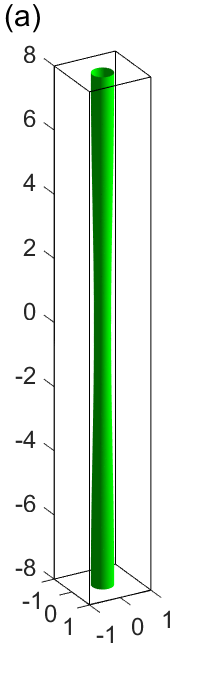}}
  \subfigure{ \includegraphics[scale=.32]{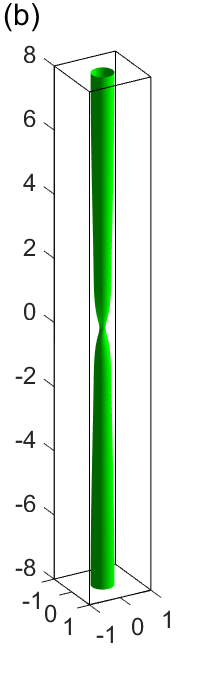}}
  \subfigure{ \includegraphics[scale=.32]{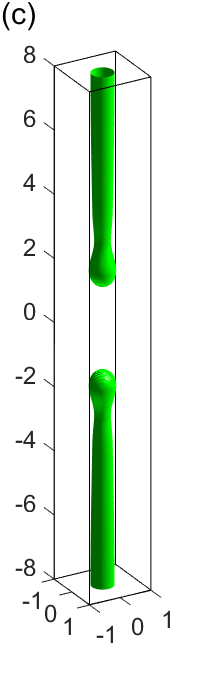}}
  \subfigure{ \includegraphics[scale=.32]{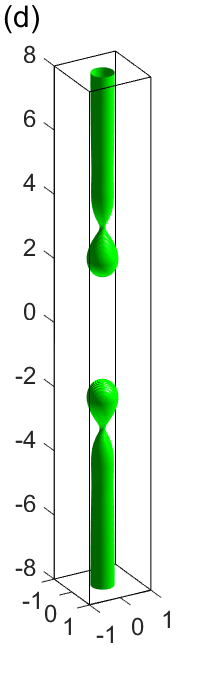}}
  \subfigure{ \includegraphics[scale=.32]{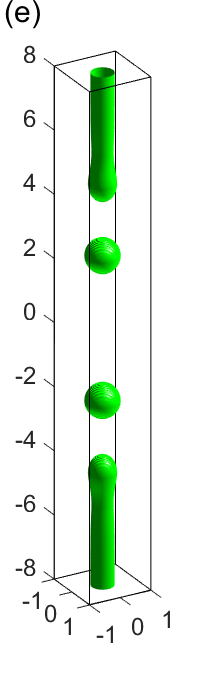}}
  \subfigure{ \includegraphics[scale=.32]{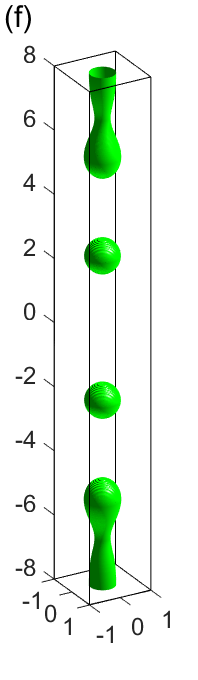}}
  \subfigure{ \includegraphics[scale=.32]{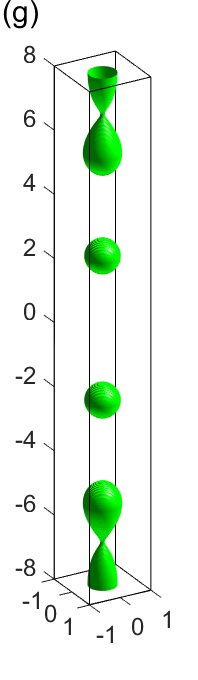}}
  \subfigure{ \includegraphics[scale=.32]{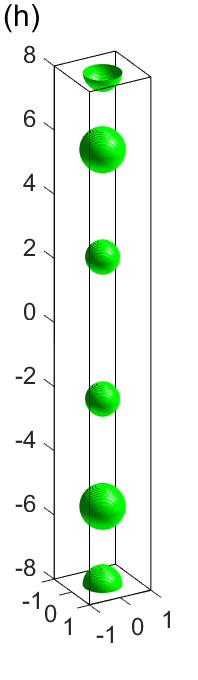}}\\
  \subfigure{ \includegraphics[scale=.32]{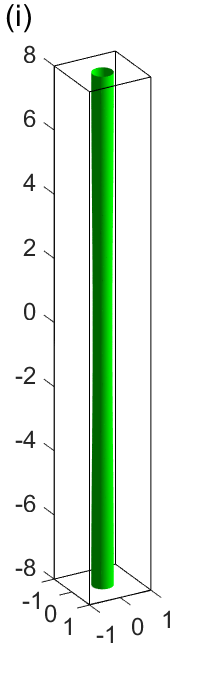}}
  \subfigure{ \includegraphics[scale=.32]{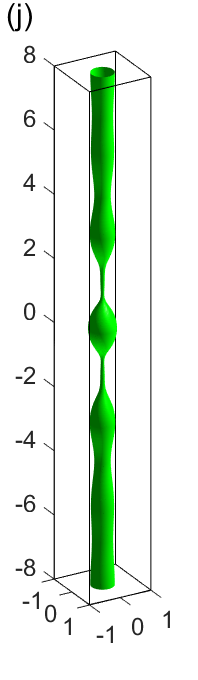}}
  \subfigure{ \includegraphics[scale=.32]{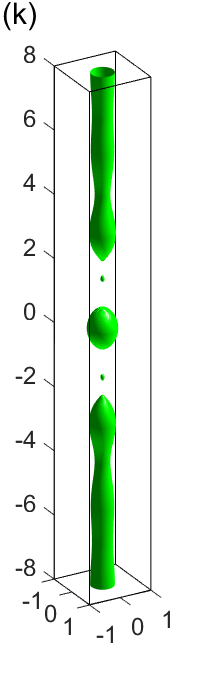}}
  \subfigure{ \includegraphics[scale=.32]{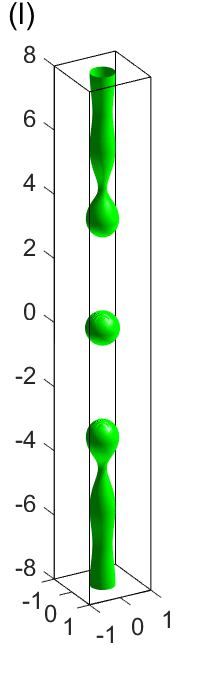}}
  \subfigure{ \includegraphics[scale=.32]{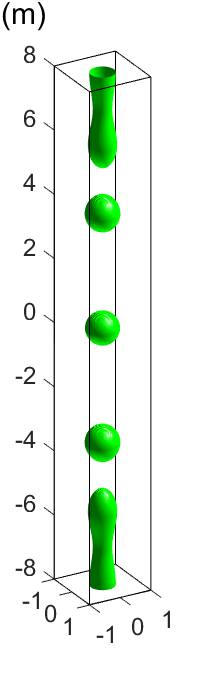}}
  \subfigure{ \includegraphics[scale=.32]{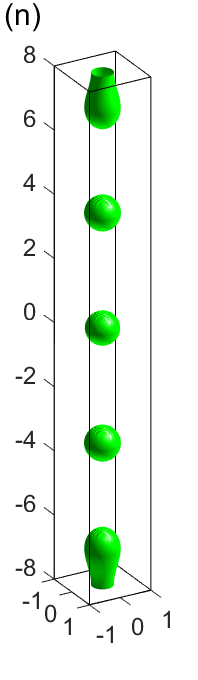}}
  \subfigure{ \includegraphics[scale=.32]{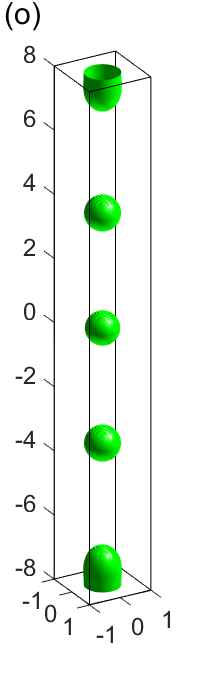}}
  \subfigure{ \includegraphics[scale=.32]{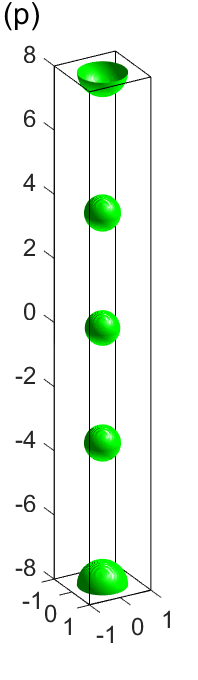}}\\
 \caption{Two sequences of multiple breakup events, one in the diffusion-dominated regime (a)-(h)
 and the other in the Stokes regime (i)-(p):
 (a)-(h)
 for $\bar{H}=8$, $\bar{\xi}=0.02$, $B=10$ and the initial interfacial profile with
 $\bar{r}_0(0)=0.28$ and $\bar{r}_0(\pm \bar{H})=0.32$,
 and the eight snapshots are taken at $\bar{t}=0.8$, $36$, $52$, $64.8$, $76$, $112$, $120.8$, and $160$;
 (i)-(p)
  for $\bar{H}=8$, $\bar{\xi}=0.02$, $B=0.0005$ and the initial interfacial profile with
 $\bar{r}_0(0)=0.28$ and $\bar{r}_0(\pm \bar{H})=0.32$,
 and the eight snapshots are taken at $\bar{t}=0.008$, $1.92$, $1.96$, $2.16$, $2.28$, $2.68$, $2.88$, and $4.00$.
 }\label{fig:multiple_drops_diffusion}
 \end{figure}

Figure \ref{fig:multiple_drops_diffusion} shows a sequence of
multiple breakup events in the diffusion-dominated regime.
With the initial interfacial profile given by $r_0(z)$, the first pinch-off starts at $z=0$.
Driven by the chemical potential gradient, the two tip ends produced from the first pinch-off
quickly retract toward $z=\pm H$ respectively. This generates a thickening head followed by
a thinning neck in each of the two half spaces (with the reflection symmetry about $z=0$).
The growing head eventually detaches from the long liquid thread at the continuously thinning neck.
This produces a droplet that quickly relaxes to a spherical shape and a new tip end that
undergoes a quick retraction once again. The droplet production cycle will be repeated if
the liquid thread left from the previous cycle is long enough.
Characterized by the ``end-pinching'' mode of breakup,
this diffusion-driven process is phenomenologically similar to the periodic mass shedding of
a retracting solid film step \cite{MASS_SHEDDING2000}.

For comparison purpose, figure \ref{fig:multiple_drops_diffusion} also shows a sequence of
multiple breakup events in the Stokes regime.
Although the initial interfacial profile is still given by $r_0(z)$, no pinch-off occurs
at $z=0$ this time. It is observed that an instability mode of several periods is developed
in the initial stage, with the number of periods determined by the ratio of $2H$ to $2\pi R$.
Given the reflection symmetry about $z=0$, whether a pinch-off occurs at $z=0$ or not
depends on whether the instability mode presents a trough or a crest there.
The growth of the instability mode eventually leads to the production of
the first three mother drops with small and transient satellites in between.
It is important to note that
out of the instability mode with an approximate periodicity in the axial direction,
the first three mother drops are produced in a short time window.
Also seen in figure \ref{fig:mother_satellite_flow}, this feature is
distinct from the droplet production in the diffusion-dominated regime where droplets
are produced one after another in a sequential order at a retracting tip end.
Note that when the first three drops are produced, the rest of the liquid thread becomes
the fourth and last mother drop, which finally reaches a spherical shape centered at $z=H$,
which is equivalent to $z=-H$ due to the periodic boundary conditions.

\begin{figure}[!htbp]
 \centering
\subfigure{ \includegraphics[scale=.32]{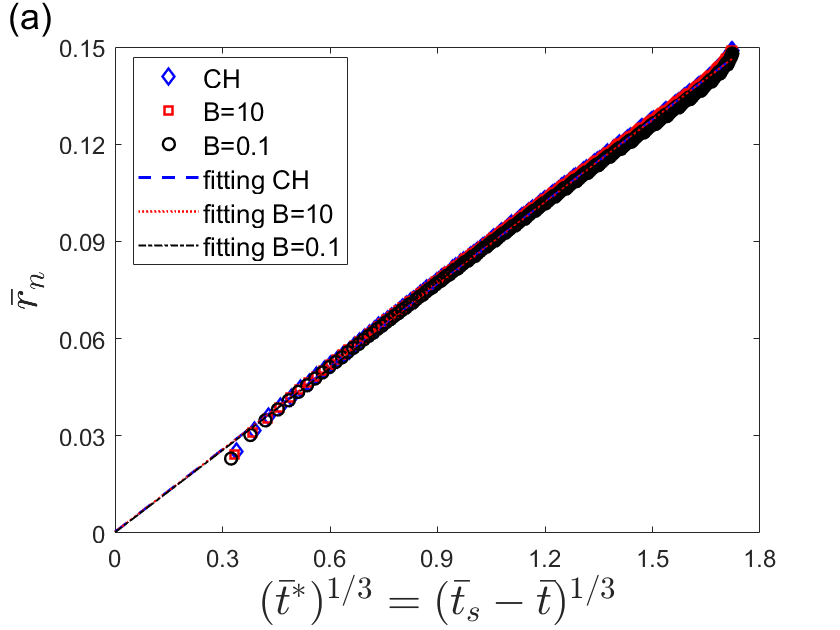}}
\subfigure{ \includegraphics[scale=.32]{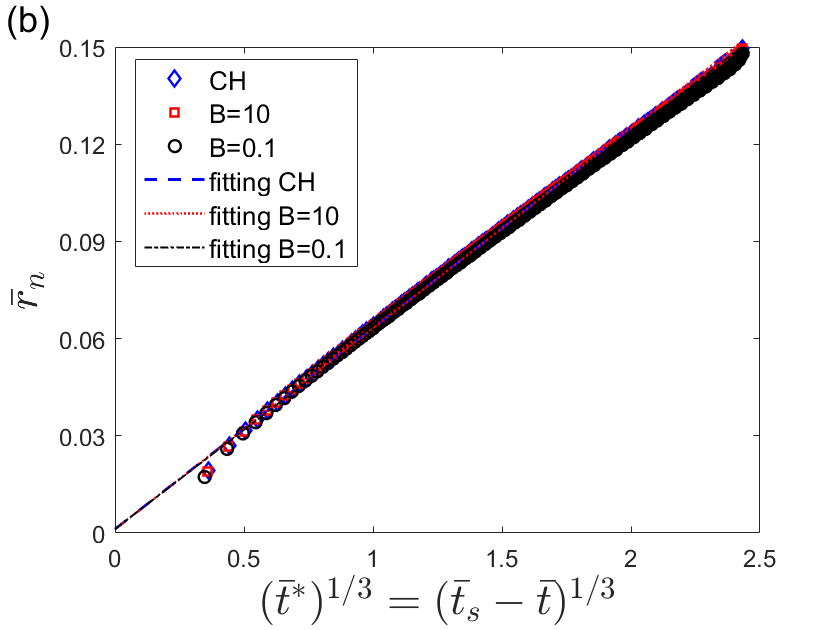}}
 \caption{The dimensionless neck radius $\bar{r}_n$ plotted as a function of $\left(\bar{t}^*\right)^{1/3}$,
 with $\bar{t}^*=\bar{t_s}-\bar{t}$ being the dimensionless time to pinch-off.
 The data are obtained by using $B=0.1$ and $B=10$ and
 by solving the CH equation with $\bar{\mathbf{v}}=0$ (corresponding to $B=\infty$):
 (a) for $\bar{H}=2$, $\bar{\xi}=0.005$, and the initial interfacial profile with
 $\bar{r}_0(0)=0.15$ and $\bar{r}_0(\pm \bar{H})=0.5$, and
 (b) is the same as (a), but with $\bar{\xi}=0.0025$.
 }\label{fig:neck_radius_diffusion}
 \end{figure}

By making the pinch-off occur at $z=0$ about which the evolving interface is symmetric,
the neck radius $r_n$ of the liquid thread is measured as a function of time at $z=0$.
As shown in figure \ref{fig:neck_radius_diffusion}, the time dependence of $\bar{r}_n$
close to the pinch-off gives $\bar{r}_n \propto \left(\bar{t}^*\right)^{1/3}$,
as predicted above.

\begin{figure}[!htbp]
 \centering
{ \includegraphics[scale=.4]{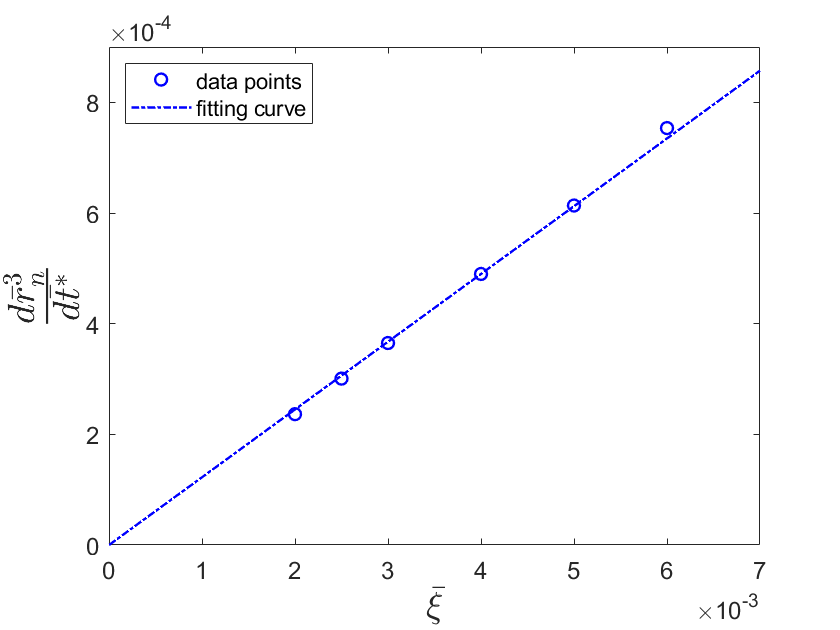}}
 \caption{The slope determined from $\frac{d \bar{r}_n^3}{d \bar{t}^*}$ vs $\bar{\xi}$ for $\bar{\xi}\le 0.006$.
 The data are obtained by solving the CH equation with $\bar{\mathbf{v}}=0$ (corresponding to $B=\infty$)
 for $\bar{H}=2$ and the initial interfacial profile with
 $\bar{r}_0(0)=0.15$ and $\bar{r}_0(\pm \bar{H})=0.5$.
 }\label{fig:neck_radius_diffusion_slope}
 \end{figure}

In the beginning of this section, it has been shown that driven by the gradient of chemical potential,
diffusion may lead to an interfacial velocity of a magnitude
$\sim V_d=\frac{M\gamma}{\phi_0^2 r_n^2}$.
Combining this estimation with $\frac{dr_n}{dt}\propto V_d$, we obtain
$\frac{dr_n^3}{dt}\propto \frac{M\gamma}{\phi_0^2 }$, which gives the dimensionless relation
$\frac{d \bar{r}_n^3}{d \bar{t}^*}\propto \bar{\xi}$.
Figure \ref{fig:neck_radius_diffusion_slope} shows that the slope obtained from
$\frac{d \bar{r}_n^3}{d \bar{t}^*}$ is proportional to $\bar{\xi}$ for $\bar{\xi}\le 0.006$.
Here we add that if the value of $B$ is large enough, then the numerical results
obtained by solving the CHNS system (\ref{CHNS}) are indistinguishable from those
obtained by solving the CH equation (\ref{CHNS1}) with $\bar {\mathbf{v} }=0$.
This means that the viscous flow is effectively suppressed by the high viscosity, and hence
the interfacial motion is driven by the diffusive transport.

\subsection{\label{sec:crossover} Crossover due to the competition between advection and diffusion}

\begin{figure}[!htbp]
 \centering
  \subfigure{ \includegraphics[scale=.35]{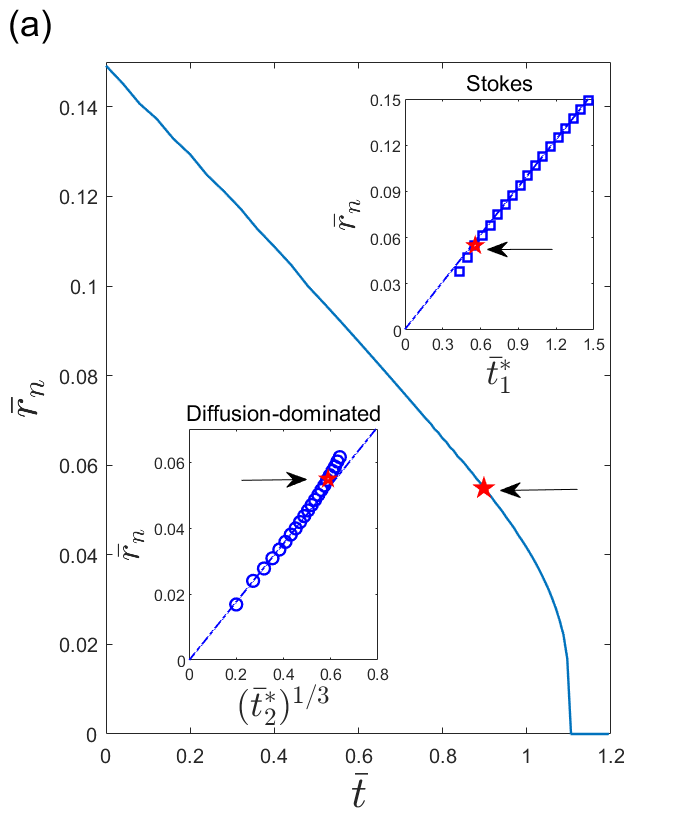}}
  \subfigure{ \includegraphics[scale=.35]{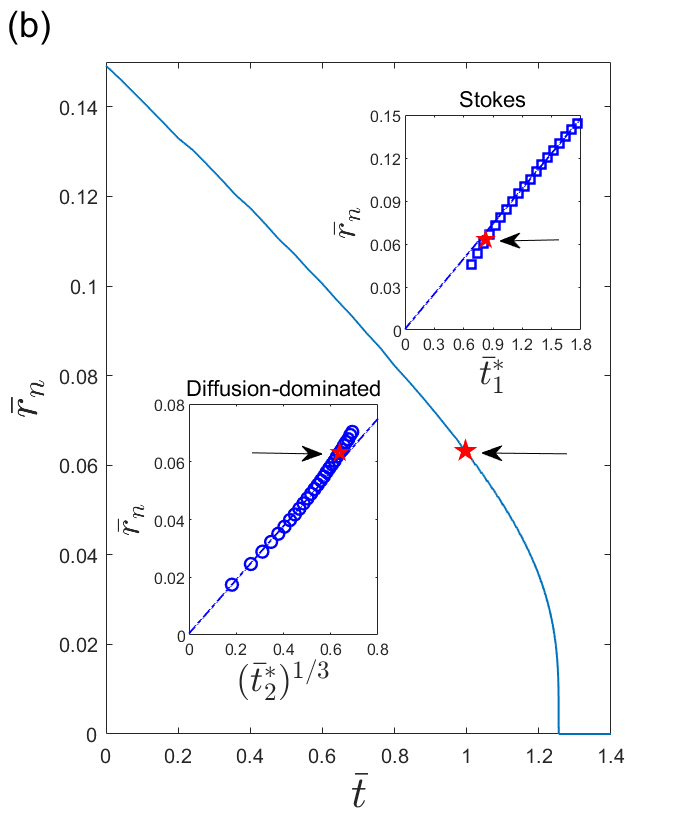}}
 \caption{Time dependence of $\bar{r}_n$ showing two different scaling behaviors
 in the initial and final stages respectively, with the crossover occurring around $r_n\approx 2l_c$:
 (a) for $\bar{H}=2$, $\bar{\xi}=0.005$, $B=0.0015$, and the initial interfacial profile with
 $\bar{r}_0(0)=0.15$ and $\bar{r}_0(\pm \bar{H})=0.8$. (b) is the same as (a), but with $B=0.002$.
 The insets shows the Stokes regime (upper right) and the diffusion-dominated regime (lower left).
 The point of $r_n=2l_c$ (i.e., $\bar{r}_n=\sqrt{2B}$) is marked to indicate the occurrence of crossover.
 Each inset uses its own time of pinch-off $t_s$ to define $\bar{t}^*$(i.e., $\bar{t}^*_1$ or $\bar{t}^*_2$) for a linear fitting.
 }\label{fig:crossover_radius}
 \end{figure}

In the time evolution of an interface, the crossover from the Stokes regime to
the diffusion-dominated regime is observable if the decreasing neck radius $r_n(t)$
meets the characteristic length $l_c$ at a certain time instant.
Given the limited accessible range of $\bar {r}_n$ in our simulations, we need to choose
a particular value of $B$ so as to numerically observe the crossover.
In figure \ref{fig:crossover_radius}, the time dependence of the neck radius is plotted to show
that two distinct scaling behaviors can be observed in the initial and final stages respectively,
with the crossover occurring around $r_n\approx 2l_c$.
Here the pinch-off occurs at $z = 0$ about which the evolving interface is symmetric.
Limited by the computational cost, the two scaling regimes shown in
each of figure \ref{fig:crossover_radius}(a) and (b) are narrow compared to those presented in
Sec. \ref{sec:Stokes_regime} and Sec. \ref{sec:diffusion_regime}.

\begin{figure}[!htbp]
 \centering
  \subfigure{ \includegraphics[scale=.23]{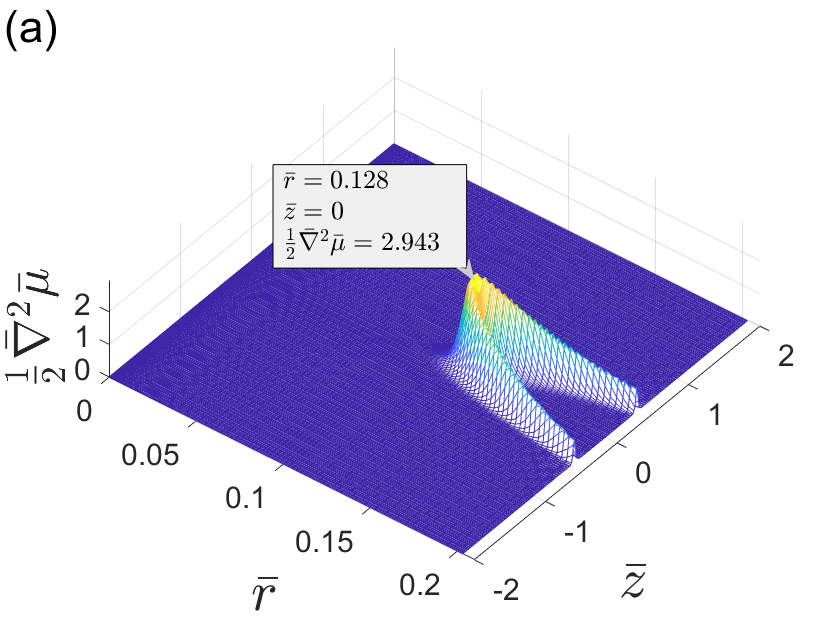}}
  \subfigure{ \includegraphics[scale=.23]{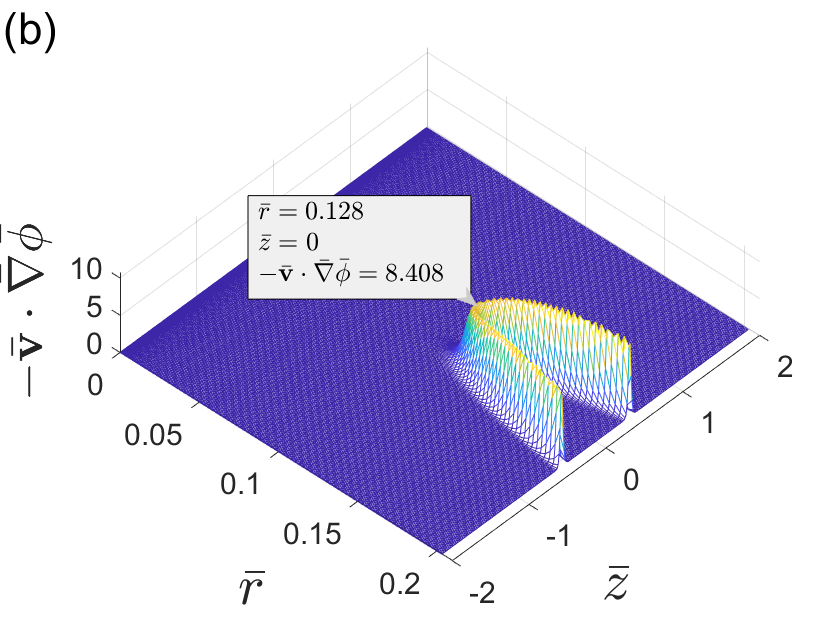}}
  \subfigure{ \includegraphics[scale=.23]{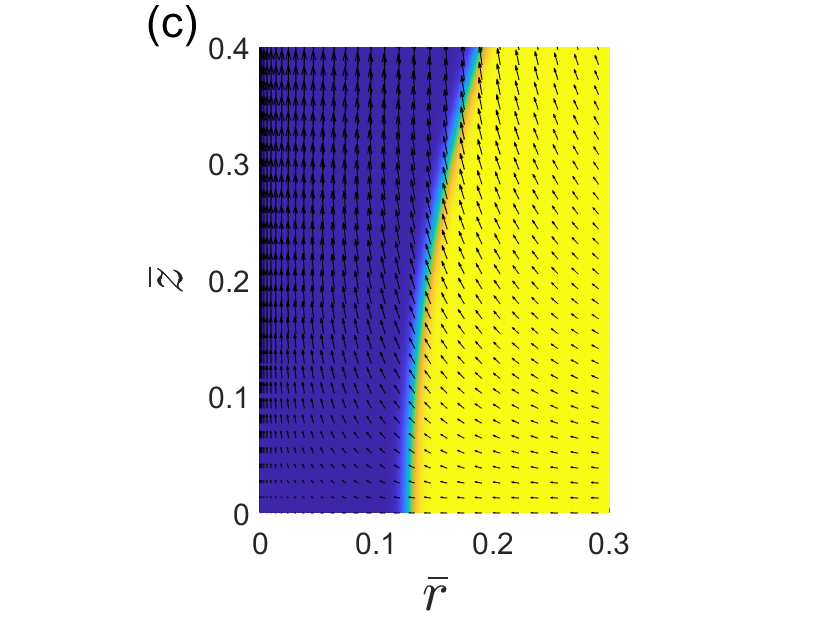}}\\
  \subfigure{ \includegraphics[scale=.23]{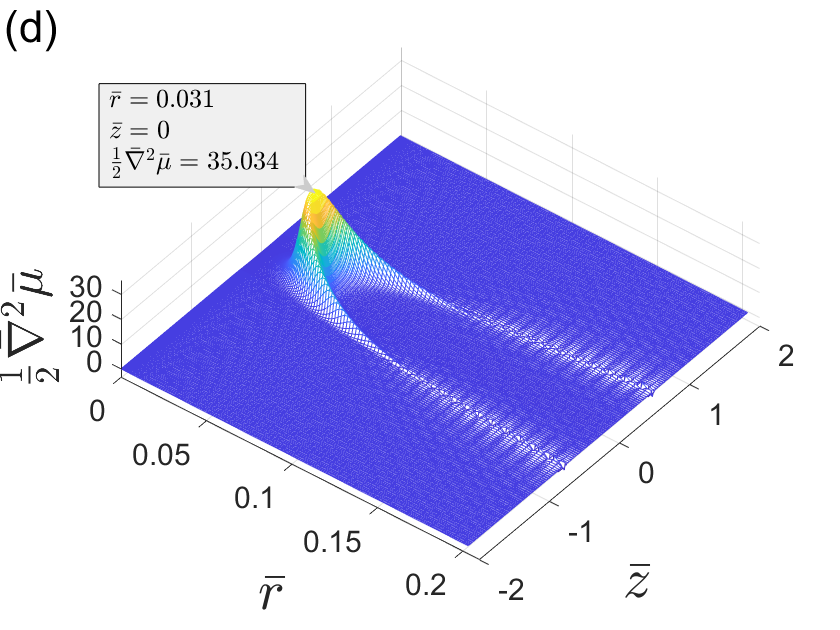}}
  \subfigure{ \includegraphics[scale=.23]{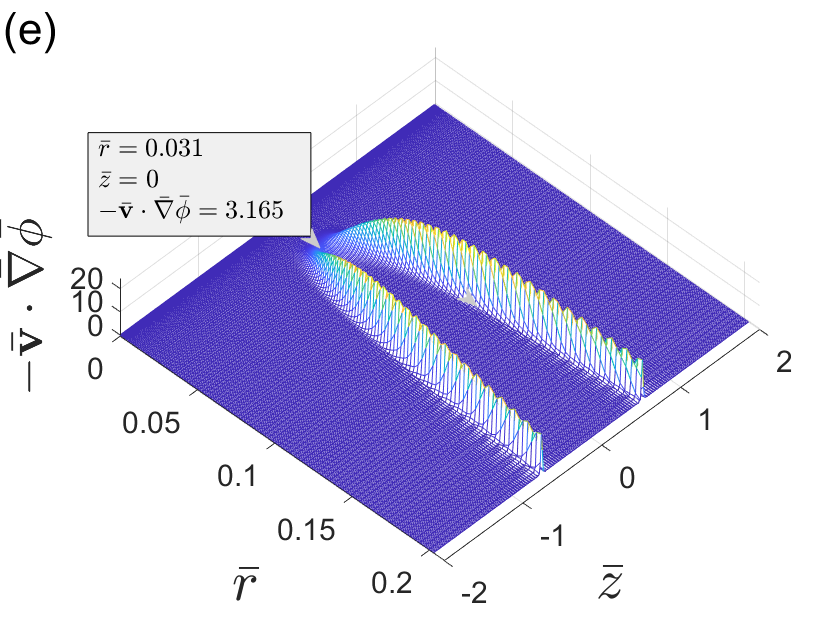}}
  \subfigure{ \includegraphics[scale=.23]{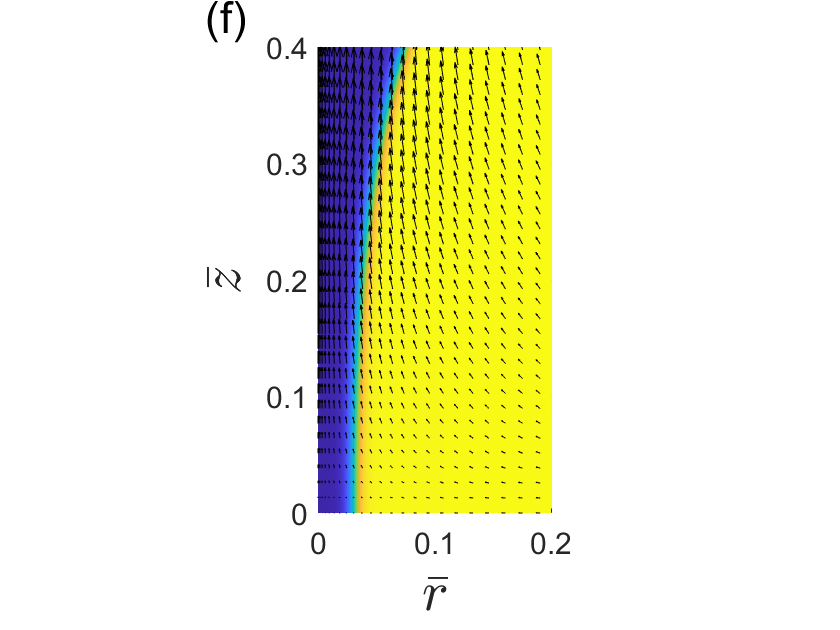}}\\
 \caption{Contributions of advection and diffusion to the evolution of interface,
 obtained for the case in figure \ref{fig:crossover_radius}(b) with $B=0.002$
 (i.e., $\frac{l_c}{L}=0.0316$):
 (a)-(c) are taken at $\bar{r}_n=0.128$ in the Stokes regime, showing the spatial distributions
 of $-\bar {\mathbf{v} } \cdot \bar{\nabla} \bar \phi$ in (a),
 $\frac{1}{2} \bar{\nabla}^2 \bar \mu$ in (b),
 and $\mathbf{v}$ (with the interface indicated by the color field) in (c), and
 (d)-(f) are taken at $\bar{r}_n=0.031$ in the diffusion-dominated regime, showing those
 corresponding to (a), (b), and (c).
 }\label{fig:crossover_advection_diffusion}
 \end{figure}

To confirm the dominance of viscous flow or diffusion in each regime, the contributions of
advection and diffusion to the evolution of interface are plotted
in figure \ref{fig:crossover_advection_diffusion}.
The key observations made for equation (\ref{CHNS1}) are as follows.
In the initial Stokes regime, $\frac{\partial \bar \phi}{\partial \bar t}$ is dominated by
the advection term $-\bar {\mathbf{v} } \cdot \bar{\nabla} \bar \phi$ whose distribution
is concentrated in the interfacial region.
The distribution of the diffusion term $\frac{1}{2} \bar{\nabla}^2 \bar \mu$ is concentrated
in the interfacial region close to the neck,
displaying a magnitude that is much smaller than that of the advection term there.
It is therefore confirmed that the interfacial evolution is predominantly driven by the flow.
In the final diffusion-dominated regime, the physical picture is very different.
The distribution of the advection term is still concentrated in the interfacial region,
with its magnitude being suppressed close to the neck.
The distribution of the diffusion term is still concentrated in the interfacial region close to the neck,
displaying a magnitude that is much larger than that of the advection term there.
These numerical results confirm that close to the neck, the interfacial evolution is
predominantly driven by the diffusion. Away from the neck, the interfacial curvature gets smaller,
and the flow gradually takes over from the diffusion at larger length scale.
A direct comparison between the flow fields in figures \ref{fig:crossover_advection_diffusion}(c) and (f)
shows that as the neck radius approaches $l_c$, the flow close to the neck is suppressed.

\begin{figure}[!htbp]
 \centering
  \subfigure{ \includegraphics[scale=.3]{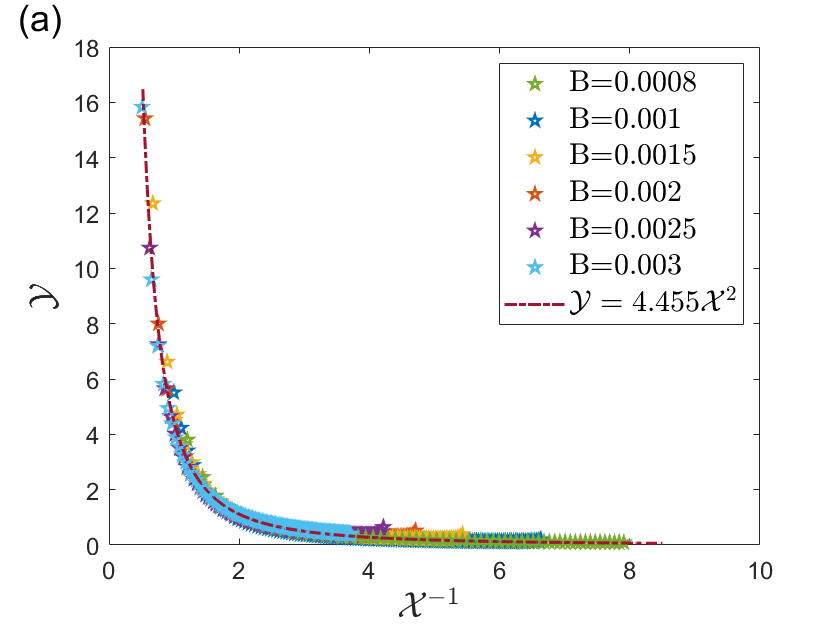}}
  \subfigure{ \includegraphics[scale=.3]{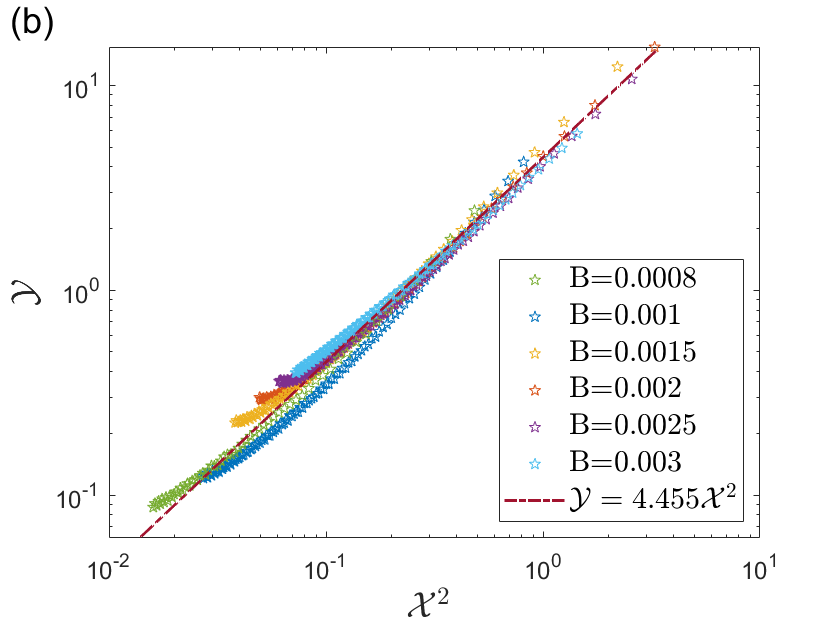}}\\
 \caption{${\cal Y}$ vs ${\cal X}$ plotted in two different ways,
 where ${\cal Y}$ measures the ratio of diffusion to advection at the neck
 as defined in the main text, and ${\cal X}$ is defined by
 ${\cal X}=\frac{l_c}{r_n}=\frac{1}{\bar{r}_n}\sqrt{\frac{B}{2}}$.
 Here $\bar{H}=1.5$, $\bar{\xi}=0.005$, and the initial interfacial profile with
 $\bar{r}_0(0)=0.125$ and $\bar{r}_0(\pm \bar{H})=0.85$.
 Different values of $B$ lead to different ranges of ${\cal X}$ measured from
 the evolving interfaces, and data collected from different evolution processes
 all follow ${\cal Y}=4.455{\cal X}^2$ approximately,
 showing semi-quantitative agreement with equation (\ref{eq:V_d-V_c}).
 }\label{fig:crossover_l_c/r_n}
 \end{figure}

The competition between diffusion and viscous flow is controlled by the characteristic length $l_c$
measured relative to the neck radius $r_n$, as expressed by equation (\ref{eq:V_d-V_c}).
A series of simulations has been carried out for a quantitative verification of this dependence
over a wide range of $\frac{l_c}{r_n}$.
Figure \ref{fig:crossover_l_c/r_n} shows ${\cal Y}$ vs ${\cal X}$ obtained from different
evolution processes with different values of $B$ (i.e., different values of $\frac{l_c}{L}$).
Here ${\cal X}$ is a dimensionless time-dependent variable, defined by ${\cal X}=\frac{l_c}{r_n}$,
and ${\cal Y}$ measures the ratio of diffusion to advection, defined by
$${\cal Y}=\frac
{\max\left(\left| \frac{1}{2} \bar{\nabla}^2 \bar \mu \right|\right)}
{\max\left(\left| -\bar {\mathbf{v} } \cdot \bar{\nabla} \bar \phi \right|\right)},$$
in which $\max\left(\left| -\bar {\mathbf{v} } \cdot \bar{\nabla} \bar \phi \right|\right)$
and $\max\left(\left| \frac{1}{2} \bar{\nabla}^2 \bar \mu \right|\right)$ are taken
in the interfacial region at $z=0$ where $r_n$ is measured.
Note that different values of $B$ lead to different ranges of ${\cal X}$ measured from
the evolving interfaces. Figure \ref{fig:crossover_l_c/r_n} presents the data collected from
different evolution processes, all of which follow ${\cal Y}=4.455{\cal X}^2$ approximately
and hence agree with equation (\ref{eq:V_d-V_c}) semi-quantitatively.

\section{\label{sec:conclusion} Concluding remarks}

Derived by applying Onsager's variational principle, the CHNS model has been employed
to investigate the pinch-off dynamics of a liquid thread. Numerical simulations have been performed
in a cylindrical domain with axisymmetry, and a characteristic length scale $l_c$ has been introduced to
measure the competition between diffusion and viscous flow in interfacial motion.
This length scale is adjustable in the model and can approach micrometer scale
for aqueous two-phase systems close to the critical point.
Consider a pinch-off process in which the pinching neck has the neck radius $r_n(t)$.
When $ r_n \gg l_c$, the pinch-off dynamics is in the Stokes regime where the interfacial motion is
driven by viscous flow. We have numerically verified that the neck radius is linearly dependent on time,
with a dimensionless prefactor close to that obtained in literature \cite{Nagel1999,Lister1998,Lister1999},
i.e., $r_n(t)=0.0335\frac{\gamma}{\eta}(t_s-t)$.
On the other hand, when $r_n \ll l_c$, the pinch-off dynamics is in the diffusion-dominated regime
where the interfacial motion is driven by bulk diffusion. We have analytically derived and
numerically verified the power-law relationship between the neck radius and the time to pinch off,
i.e., $r_n(t) \sim (t_s-t)^{1/3}$,
which agrees with the recent experimental observation in aqueous two-phase systems \cite{Xu2019}.
Finally, when $r_n \approx 2l_c$ during the pinch-off process, the crossover from the Stokes regime to
the diffusion-dominated regime occurs. Our numerical results have provided strong evidence to confirm
the relation $\displaystyle\frac{V_d}{V_\eta}\approx \displaystyle\frac{l_c^2}{r_n^2}$,
which is derived in our theoretical analysis.
Besides the effects of the characteristic length scale $l_c$, we have also presented numerical examples
for the breakup of long liquid filaments. Qualitatively different phenomena have been observed
in different scaling regimes with accompanying physical interpretation.

We understand that the present study has limitations.
Firstly, we have only simulated our model in a bounded domain
with the interfacial thickness $\xi$ being not too small
due to the limited computational capability, while in the ideal situation, the pinch-off dynamics,
which admits similarity solutions, should be simulated in an infinite domain
with $\xi$ being sufficiently small.
Secondly, our numerical simulations have only treated the simplest case in which
the two fluid phases separated by the interface have equal density, equal viscosity,
and equal mobility (diffusion coefficient).
Actually, different self-similar profiles can be developed in the pinching neck
for different viscosity ratios in the Stokes regime \cite{Lister1999}.
Furthermore, in the recent experiment \cite{Xu2019},
components of different diffusion coefficients are involved, and
the outer phase (surrounding phase) has different viscosity from that of
the inner phase (liquid thread).
Nevertheless, the primary purpose of the present work is to show that the CHNS model
can act as a minimal model to demonstrate the presence of two distinct scaling regimes,
i.e., the Stokes regime and the diffusion-dominated regime, and the crossover between them.
The quantitative effects of the viscosity ratio and mobility ratio can be investigated
by using $\phi$-dependent viscosity and mobility in the CHNS model
\cite{ding2007,dong2012,kim2005} or some more sophisticated models.
With a more accurate and efficient numerical scheme, systematic simulations can be
carried out to remove the above limitations.


\begin{acknowledgments}
The work of F. Huang and W. Bao was supported by the Ministry of Education of Singapore grant MOE2019-T2-1-063 (R-146-000-296-112),
and the work of T. Qian was supported by the Hong Kong RGC grants CRF No. C1006-20WF and GRF No. 16306121.
\end{acknowledgments}

\begin{appendix}
\section{\label{sec:Onsager} Onsager's variational principle}

The thermodynamic state of a closed system can be described by a set of coarse grained state variables
$\alpha_i$ $(i=1,\cdot\cdot\cdot,n)$, which are measured relative to their equilibrium values.
The entropy function of the system $S$ has a maximum at equilibrium, denoted by $S_e$,
and the difference $\Delta S=S-S_e$ may be written as
$\Delta S=-\displaystyle\frac{1}{2}\sum_{i,j=1}^{n}\beta_{ij}\alpha_i\alpha_j$ in an expansion to the quadratic order,
where the coefficient matrix $\beta$ is symmetric and positive definite.
Out of equilibrium, the thermodynamic force conjugate to $\alpha_i$ is defined by
$$ X_i=\displaystyle\frac{\partial }{\partial \alpha_i} \Delta S ,$$
which is a linear in $\{\alpha_j\}$ for small perturbations away from equilibrium.
Physically, these forces are to bring the system back to equilibrium.
The linear irreversible thermodynamics \cite{De_Groot} states that a macroscopic state
evolves in time according to the linear equations
$$\displaystyle\frac{d}{dt}\alpha_i (t)=\sum_{j=1}^{n}L_{ij} X_j (t),$$
where the phenomenological kinetic coefficients $L_{ij}$ satisfy the reciprocal symmetry
$L_{ij}=L_{ji}$, which was derived from the microscopic reversibility in the original works
by Onsager \cite{Onsager1931a, Onsager1931b}. It is this reciprocal symmetry that enables us
to formulate a variational principle governing the time evolution of the system \cite{Onsager1931a, Onsager1931b}.

Originally formulated for isolated systems, Onsager's variational principle can be
readily extended to a description of isothermal systems with constant temperature
being maintained in space and time \cite{Qian2006, Doi2011, Doi2013}.
Let the rates of change of the state variables $\alpha_i$ be denoted by $\dot{\alpha}_i$.
An action function, commonly referred to as the Rayleighian \cite{Doi2011}
and denoted by ${\cal{R}}$, can be put in the form of
$${\cal{R}}=\displaystyle\frac{1}{2}\sum_{i,j=1}^{n}\zeta_{ij}\dot{\alpha}_i\dot{\alpha}_j
+\sum_{i=1}^{n}\displaystyle\frac{\partial F}{\partial \alpha_i}\dot{\alpha}_i,$$
which includes two physically distinct terms on the right-hand side.
The first term is the dissipation function, hereafter denoted by $\Phi$,
which is defined as half the rate of free energy dissipation.
Note that $\Phi$ is quadratic in $\{ \dot{\alpha}_i \}$, and
the friction coefficient matrix $\zeta$ is symmetric and positive definite.
The second term is the rate of change of the free energy $F=F(\alpha_1,\cdot\cdot\cdot,\alpha_n)$,
hereafter denoted by $\dot{F}$, which is linear in $\{ \dot{\alpha}_i \}$.
Minimizing $\cal{R}$ with respect to the rates $\dot{\alpha}_i$ $(i=1,\cdot\cdot\cdot,n)$, we obtain
$$\sum_{j=1}^{n}\zeta_{ij}\dot{\alpha}_j= -\displaystyle\frac{\partial F}{\partial \alpha_i}.$$
Expressing the balance between reversible and dissipative forces, these equations
govern the time evolution of the state variables.
Furthermore, as a result of variation, the rate of free energy dissipation
$\dot{F}=\sum_{i=1}^{n}\left({\partial F}/{\partial \alpha_i}\right)\dot{\alpha}_i$
equals $-2\Phi$, as required by thermodynamic consistency \cite{Xu2015}.

\section{\label{sec:numerical} CHNS in 3D with axisymmetry and the numerical scheme}

To reduce the computational cost in our simulations, we first rewrite the 3D problem (\ref{CHNS})
with axisymmetry as a 2D problem. In the following, we use the cylindrical coordinates $(r,\theta,z)$
and denote the corresponding velocity in the cylindrical coordinate system by
$(u,v,w)^{T}$.
We still use $\phi$ and $\mu$ to denote the phase-field variable and chemical potential respectively. With the axisymmetry,
all variables are spatially dependent on two coordinates $(r,z)$ only.
Given the initial condition $\mathbf{v}(\cdot, 0)=0$ and the boundary conditions described in Sec. \ref{sec:dimensionless},
we have $v \equiv0$. Now we have the following 2D problem in $\hat{\Omega}=\{(r,z): 0<r<1, -\bar{H}<z<\bar{H}\}$:
\begin{subequations}\label{CHNS_C}
\begin{align}
& \frac{\partial \phi}{\partial  t}+u \frac{\partial \phi}{\partial r}+w \frac{\partial \phi}{\partial z} =\frac{1}{2} \Big(\frac{1}{r}\frac{\partial}{\partial r}(r\frac{\partial \mu}{\partial r})+\frac{\partial^2 \mu}{\partial z^2}\Big),\label{CHNS1_C}\\
&  \mu=-\xi^2 \Big(\frac{1}{r}\frac{\partial}{\partial r}(r\frac{\partial \phi}{\partial r})+\frac{\partial^2 \phi}{\partial z^2}\Big)- \phi+\phi^3,\label{CHNS2} \\
& A B \Big(\frac{\partial  u}{\partial t}+u\frac{\partial u}{\partial r}+w \frac{\partial u}{\partial z}\Big)=-\frac{\partial p}{\partial r}+B\Big(\frac{1}{r}\frac{\partial}{\partial r}(r\frac{\partial u}{\partial r})-\frac{u}{r^2}+\frac{\partial^2 u}{\partial z^2} \Big)+\mu \frac{\partial \phi}{\partial r}, \\
& A B \Big(\frac{\partial  w}{\partial t}+u\frac{\partial w}{\partial r}+w \frac{\partial w}{\partial z}\Big)
=-\frac{\partial p}{\partial z}+B\Big(\frac{1}{r}\frac{\partial}{\partial r}(r\frac{\partial w}{\partial r})+\frac{\partial^2 w}{\partial z^2} \Big)+\mu \frac{\partial \phi}{\partial z}, \\
& \frac{1}{r}\frac{\partial}{\partial r}(r u)+\frac{\partial w}{\partial z}=0,
\end{align}
\end{subequations}
with the boundary conditions:
\begin{subequations}
\begin{align}
& \partial_r \phi(0,z)=\partial_r \phi(1,z)=\partial_r \mu(0,z)=\partial_r \mu(1,z)=\partial_r w(0,z)=0,\\
& u(0,z)=u(1,z)=w(1,z)=0.
\end{align}
\end{subequations}
Periodic boundary conditions are applied to $\phi$, $\mu$, $u$, and $w$ on $z=\pm \bar{H}$.

For the time-discretization, we adopt the second order consistent splitting scheme proposed in \cite{Shen03}:
 \begin{subequations}\label{scheme}
\begin{align}
& \frac{3 \phi^{n+1}-4\phi^n+\phi^{n-1}}{2 \tau }+\hat{u}^n \frac{\partial \hat{\phi}^n }{\partial r}+\hat{w}^n \frac{\partial \hat{\phi}^n}{\partial z}=\frac{1}{2} \Big(\frac{1}{r}\frac{\partial}{\partial r}(r \frac{\partial \mu^{n+1}}{\partial r})+\frac{\partial^2 \mu^{n+1}}{\partial z^2}\Big) \label{scheme1},\\
&  \mu^{n+1}=-\xi^2 \Big(\frac{1}{r}\frac{\partial}{\partial r}(r \frac{\partial \phi^{n+1}}{\partial r})+\frac{\partial^2 \phi^{n+1}}{\partial z^2}\Big)- \hat{\phi}^n+(\hat{\phi}^n)^3, \\
& A B \big(\frac{3 u^{n+1}-4 u^n+u^{n-1}}{2 \tau}+\hat{u}^{n} \frac{\partial \hat{u}^n}{\partial r}+\hat{w}^n \frac{\partial \hat{u}^n}{\partial z}\big) \nonumber\\
& =-\frac{\partial \hat{p}^n }{\partial r}+B\Big(\frac{1}{r}\frac{\partial}{\partial r} (r\frac{\partial u^{n+1}}{\partial r})-\frac{u^{n+1}}{r^2}+\frac{\partial^2 u^{n+1}}{\partial z^2} \Big)+\mu^{n+1} \frac{\partial \phi^{n+1}}{\partial r}, \label{scheme3}\\
& A B \Big(\frac{3 w^{n+1}-4 w^n+w^{n-1}}{2 \tau}+\hat{u}^n \frac{\partial \hat{w}^{n}}{\partial r}+\hat{w}^n \frac{\partial \hat{w}^n}{\partial z}\Big) \nonumber\\
& =-\frac{\partial \hat{p}^n}{\partial z}+B \Big(\frac{1}{r}\frac{\partial}{\partial r}(r \frac{\partial w^{n+1}}{\partial r})+\frac{\partial^2 w^{n+1}}{\partial z^2} \Big)+\mu^{n+1} \frac{\partial \phi^{n+1}}{\partial z}, \label{scheme4}
\end{align}
\end{subequations}
where $\hat{\phi}^n=2\phi^n-\phi^{n-1}$, and similarly for $\hat{u}^n$, $\hat{w}^n$, and $\hat{p}^{n}$. We update the pressure $p$ by solving the
following equations:
\begin{equation}
\begin{split}
& \Big(r\frac{\partial p^{n+1}}{\partial r},\, \frac{\partial q}{\partial r} \Big)+\Big(r\frac{\partial p^{n+1}}{\partial z}, \,\frac{\partial q}{\partial z} \Big)=\Big(rg_r^{n+1}-B\big(r\frac{\partial^2 w^{n+1}}{\partial r \partial z}-r\frac{\partial^2 u^{n+1}}{\partial z^2}\big), \, \frac{\partial q}{\partial r}\Big)\\
& +\Big(rg_z^{n+1}-B\big(\frac{\partial u^{n+1}}{\partial z}+\frac{\partial w^{n+1}}{\partial r}+r\frac{\partial^2 u^{n+1}}{\partial r\partial z}+r\frac{\partial^2 w^{n+1}}{\partial r^2} \big), \, \frac{\partial q}{\partial z} \Big),\quad  \forall q \in H^1(\hat{\Omega}),
\end{split}
\end{equation}
with $(\cdot, \cdot)$ denotes the inner product of $L^2(\hat{\Omega})$ and $g_r^{n+1}$, $g_z^{n+1}$ are defined as
\begin{subequations}
\begin{align}
& g_r^{n+1}=\mu^{n+1} \frac{\partial \phi^{n+1}}{\partial r}-AB\big({u}^{n+1} \frac{\partial {u}^{n+1}}{\partial r}+ {w}^{n+1} \frac{\partial {u}^{n+1}}{\partial z}\big),\\
& g_z^{n+1}=\mu^{n+1} \frac{\partial \phi^{n+1}}{\partial z}-AB\big({u}^{n+1} \frac{\partial {w}^{n+1}}{\partial r}+{w}^{n+1} \frac{\partial {w}^{n+1}}{\partial z}\big).
\end{align}
\end{subequations}
For the spatial discretization, due to the periodic boundary conditions in the $z$ direction,
we first reduce the 2D problem to a sequence of one-dimensional (1D) problems by using the Fourier expansion in the $z$ direction and then we adopt the Spectral-Galerkin method to solve the 1D problems by standard ways, with more details
given in \cite{Shen97}.

\begin{figure}[!htbp]
 \centering
  \subfigure{ \includegraphics[scale=.3]{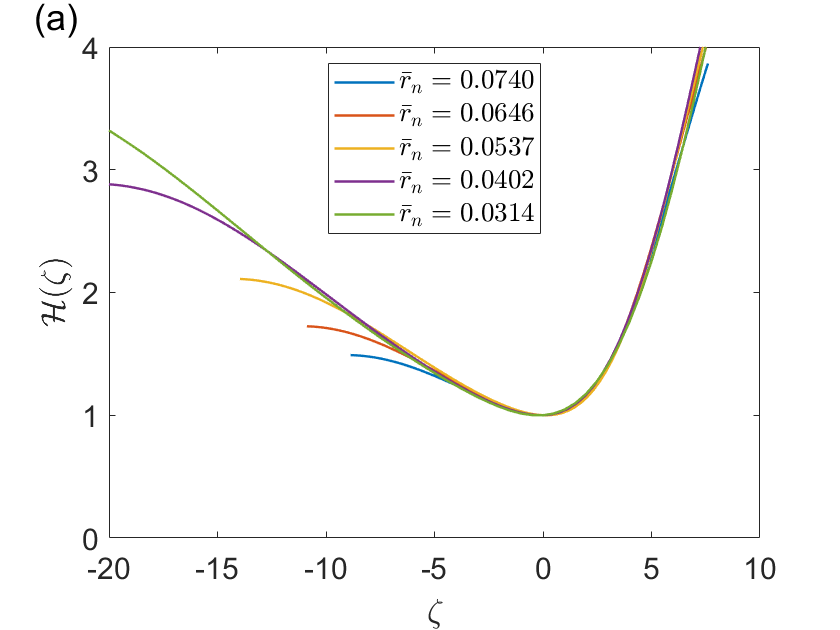}}
  \subfigure{ \includegraphics[scale=.3]{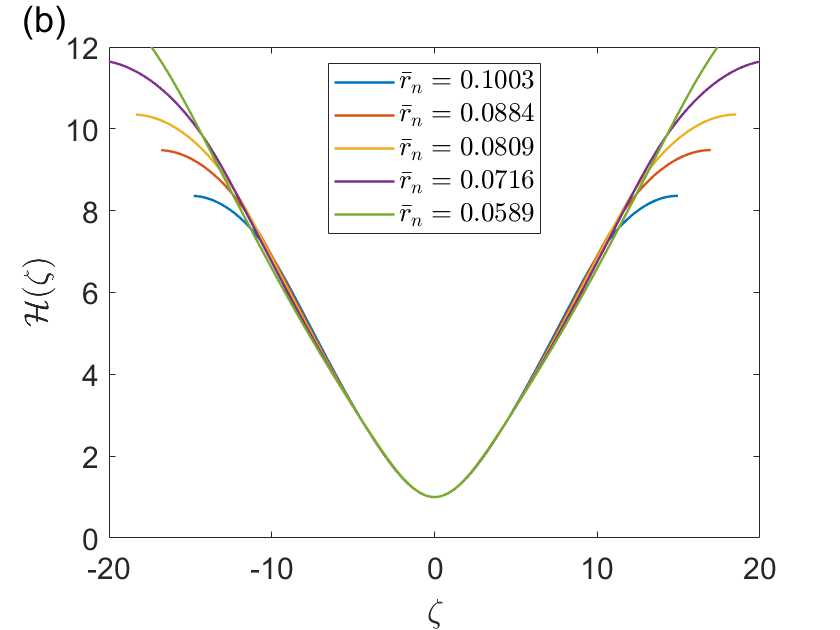}}\\
 \caption{Self-similar profiles in the pinching neck obtained at different time instants
 with different neck radius:
 (a) in the Stokes regime, computed for $\bar{H}=2$, $\bar{\xi}=0.005$, $B=0.0005$,
 and the initial interfacial profile with $\bar{r}_0(0)=0.25$ and $\bar{r}_0(\pm \bar{H})=0.35$,
 in correspondence with figure \ref{fig:neck_radius_flow_a}, and
 (b) in the diffusion-dominated regime, computed for $\bar{H}=1.5$, $\bar{\xi}=0.0025$, $B=10$,
 and the initial interfacial profile with $\bar{r}_0(0)=0.15$ and $\bar{r}_0(\pm \bar{H})=0.85$.
 }\label{fig:similarity}
 \end{figure}

The accuracy of our numerical scheme has been tested by checking the self-similarity
exhibited by the numerical solutions near pinch-off. For this purpose, we use the interfacial profile
${r}({z},{t})$ which has the minimum radius ${r}_n({t})$ occurring at ${z}={z}_n({t})$.
In a self-similar reference frame, we work with
\begin{equation}
\zeta=\frac{z-z_n(t)}{r_n(t)},
\,\,\,
{\cal H}(\zeta)=\frac{r(z,t)}{r_n(t)},
\end{equation}
with the minimum radius ${\cal H}=1$ fixed at $\zeta=0$.
Using the numerical solutions near pinch-off,
figure \ref{fig:similarity} shows self-similar profiles in the pinching neck.
Though obtained at different time instants, they overlap with each other remarkably well.
Furthermore, it is noted that the results in figure \ref{fig:similarity}(a) are
in semi-quantitative agreement with those in literatures \cite{Nagel1999,Lister1999}.
The small deviation is partly attributed to the finite radius of the cylindrical computational domain.

\end{appendix}
\newpage

\bibliographystyle{plain}
\bibliography{bibpinchoff2}

\end{document}